\documentclass[twocolumn]{aastex63}
\usepackage{natbib}
\usepackage{ulem}
\usepackage{amsmath,bm}
\usepackage{subfigure}

\begin{document}

\title{Probing the Solar Meridional Circulation using Fourier Legendre Decomposition}

\author{D.~C.~Braun}
\affiliation{NorthWest Research Associates, 3380 Mitchell Lane, Boulder, CO 80301, USA}

\author{A.~C.~Birch}
\affiliation{Max-Planck-Institut f\"{u}r Sonnensystemforschung, Justus-von-Liebig-Weg 3, 37077 G\"{o}ttingen, Germany}

\author{Y.~Fan}
\affil{National Center for Atmospheric Research, HAO Division, 
3080 Center Green Dr, Boulder, CO 80301, USA}

\correspondingauthor{D.~C.~Braun}
\email{dbraun@nwra.com}

\begin{abstract}
We apply the helioseismic methodology of Legendre Function Decomposition
to 88 months of Dopplergrams obtained by the Helioseismic and Magnetic
Imager (HMI) as the basis of inferring the depth variation of the mean meridional
flow, as averaged between 20 and 60 degrees latitude and in time, 
in both the northern and southern hemispheres.
We develop and apply control procedures
designed to assess and remove center-to-limb artifacts, using measurements obtained
by performing the analysis with respect to artificial poles at the east and west limbs. 
Forward modeling is carried out, using sensitivity functions proportional to the mode
kinetic energy density, to evaluate the consistency of the corrected 
frequency shifts with models of the depth variation of the meridional
circulation in the top half of the convection zone. The results, taken at face
value, imply substantial differences between the meridional circulation in the 
northern and southern hemisphere. The inferred presence of a return (equator-ward propagating)
flow at a depth of approximately 40 Mm below the photosphere in the northern hemisphere
is surprising and appears to be inconsistent with many other helioseismic analyses. 
This discrepancy may be the result of an inadequacy of our methodology to remove systematic
errors in HMI data. Our results appear to be at least qualitative similar to those by 
Gizon et al. (2020) which point to an anomaly in HMI data that is not present in MDI or GONG data.
\end{abstract}

\section{Introduction}

Meridional circulation is a crucial, but poorly constrained, component
of magnetic flux transport and dynamo models \citep[e.g.][]{Choudhuri1995, 
Wang1991, Wang2002, Dikpati2006, Dikpati2007}. In many types
of flux-transport models, for example, it plays an important role in determining the 
period or amplitude of the solar cycle \citep[e.g.][]{Hathaway2003,Upton2014}. 
In Babcock-Leighton type dynamo models, the meridional circulation provides
the ``conveyor belt'' that submerges the surface poloidal field of a given solar cycle
deep into the convection zone to be sheared by differential
rotation and advected towards the equator by the meridional return flow \citep[e.g.][]{Charbonneau2010}.
The meridional flow is also an important dynamical element
in theoretical and numerical models of 
solar and stellar differential rotation and convection
\citep[e.g.][]{Glatzmaier1982, Brun2002, Rempel2005, Miesch2007}.

Measurements of the surface manifestation of meridional
circulation have typically indicated poleward flows
between 10 and 20 m s$^{-1}$ \citep[e.g.][]{LaBonte1982,Topka1982,
Hathaway1996,Schou2003,Rightmire-Upton2012}. Meridional circulation is difficult
to measure accurately, since its amplitude is considerably smaller than, say,
solar rotation.  Helioseismic analyses of the subsurface properties of
this flow are particularly challenging.
The frequencies of global $p$ modes are insensitive to
first order to the meridional circulation, unlike the rotational splitting which
has facilitated the successful determination of the subsurface properties of rotation 
\citep[e.g.][]{Brown1989,Schou1998}. 
However, meridional flows have been
measured and modeled with a variety of local seismic methods.
Many studies have mostly focused on
the meridional circulation near the surface (e.g.\ within a few
tens of Mm below the surface), where the sensitivity of helioseismic
measurements to flows is highest 
\citep[e.g.][]{Gonzalez-Hernandez1999,
Haber2002, Hughes2003, Zhao2004, Gonzalez-Hernandez2008, Komm2015a, Komm2015b, Komm2018}.

Probing the deeper properties of the meridional circulation is difficult
due, in large part, to expected low signal-to-noise values 
of the measurements \citep{Braun2008} and the presence of systematic artifacts
\citep[e.g.][]{Duvall2009,Zhao2012b}. 
Nonetheless, numerous studies involving measurements and inferences of the 
deeper properties of the meridional flow have been carried out
\citep[e.g.][]{Giles1997, Braun1998b, Chou2005, Mitra-Kraev2007, Schad2013, Woodard2013, 
Zhao2013, Kholikov2014a, Kholikov2014b, Liang2015a, Rajaguru2015, Liang2018, Gizon2020}.
The last decade in particular has seen a renewal of interest in probing 
the deep meridional circulation, 
making use of the long duration datasets provided by the ground-based
Global Oscillation
Network Group (GONG) instruments \citep[][]{Harvey1996,Harvey1998} and the Helioseismic and Magnetic Imager
(HMI) instrument \citep[][]{Scherrer2012,Schou2012} onboard the {\it Solar Dynamics Observatory}
(SDO) spacecraft \citep[][]{Pesnell2012}. 
However, there is not yet a consensus on the general structure and properties of the 
meridional circulation, especially in the deeper two-thirds of the convection zone.

The purpose of this study is to perform a follow-up to the meridional-flow measurements
carried out by \cite{Braun1998b}. In that study, we applied the method now
known as Fourier Legendre decomposition (hereafter FLD) to one month of GONG data and 8 days
of Dopplergrams from the Michelson Doppler Imager (MDI) instrument \citep{Scherrer1995} 
onboard the {\it Solar and Heliospheric Observatory}. 
With various modifications, the FLD method has been employed in a number of subsequent
studies of solar meridional circulation \citep{Krieger2007,Doerr2010} as well
as undergone development and validation studies \citep{Roth2016,Hecht2018}. 
We note that the method used by \cite{Mitra-Kraev2007} is closely related to 
FLD analysis, but applied selectively to waves propagating along the central meridian.
Novel aspects of the present work include: the development of procedures to 
assess and remove the center-to-limb artifact, the application 
of FLD to a long duration (7.3 years) of HMI Dopplergrams, a comparison
and assessment of different ridge-peak finding methods,
and an assessment of the sensitivity of the measurements to azimuthal order.
We apply an updated version of the forward modeling approach employed by \cite{Braun1998b}
in order to infer the latitude-averaged meridional flow, for each hemisphere, 
in the top half of the convection zone. 

\section{Data Analysis}\label{sec.data}

\subsection{Fourier Legendre Decomposition}\label{sec.fld}

The basic concept of FLD is similar to 
ring-diagram analysis \citep{Hill1988}, and infers subsurface flows by
measuring and modeling the Doppler distortion in the power spectra.
Compared to most ring-diagram methods, the FLD technique is optimized for
the detection of meridional flows by properly accounting for
spherical geometry and using
combinations of Legendre functions of the first and
second kind to characterize poleward or equatorward propagating waves.
The basis for the technique is the expansion of the observed Dopplergram
signal $\delta V$, a function of colatitude $\theta$, azimuthal angle $\phi$
and time $t$, into traveling wave components:
\begin{multline}
\label{eqn:expansion}
\delta V (\theta ,\phi,t)= \sum _{\ell m \nu}  
e^{i(m\phi +2\pi \nu t)} [A_{\ell m \nu} \Theta_{\ell}^m (\cos \theta)\\
+ B_{\ell m \nu} (\Theta_{\ell}^m)^\ast (\cos \theta)],
\end{multline}
where $\nu$ is the temporal frequency, $\ell$ is the degree, and $m$ is the azimuthal
order.  $A_{\ell m \nu}$ and $B_{\ell m \nu}$ are the
complex amplitudes of poleward and equatorward waves respectively.
$\Theta_{\ell}^m$ is a normalized function,
\begin{equation}
\label{eqn:legendre1}
\Theta_{\ell}^m (\cos \theta) \equiv N_{\ell}^m [ P_{\ell}^m (\cos \theta)
 - \frac{2i}{\pi} Q_{\ell}^m (\cos \theta)],
\end{equation} 
where $P_{\ell}^m$ and $Q_{\ell}^m$ are Legendre functions of the
first and second kind respectively and $N_{\ell}^m$ is a normalization factor.
The expansion of waves in terms of propagating functions $\Theta_{\ell}^m$ has been utilized
for many decades in nuclear and quantum physics \citep[e.g.][]{Nussenzveig1965,Fuller1975}
where the operation is called ``nearside-farside decomposition.''
In helioseismology, the first application of this expansion was carried out by
\cite{Bogdan1993} as an extension to spherical coordinates
of Fourier Hankel decomposition \citep[e.g.][]{Braun1987,Braun1988}, and used to measure the
absorption and scattering of $p$ modes by sunspots. 

The functions $\Theta_{\ell}^m$ are orthogonal such that
\begin{equation}
\label{eqn:orthog}
\int_{\theta_1}^{\theta_2} \Theta_{\ell}^{m} (\cos \theta) (\Theta_{\ell'}^{m'})^{*} (\cos \theta) \,d(\cos \theta) = 
            \delta_{\ell \ell'} \delta_{m m'},
\end{equation}
over the annular domain $( \theta_1, \theta_2 )$, where $\delta_{i j}$
is the Kronecker delta function. 
As \cite{Bogdan1993} note, the 
condition described by equation (\ref{eqn:orthog}) generally requires non-integer values of $\ell$. 
For practical purposes, 
such as the ability to compute Legendre functions using recursion 
relations \citep{Press1992}, a restriction to integer values of degree renders equation
(\ref{eqn:orthog}) approximately true for a subset of equally spaced $\ell$. 
Furthermore, the orthogonality condition restricts the range of azimuthal order $m$
for each degree $\ell$, 
such that 
\begin{equation}
\label{eqn:impact}
\frac{|m|}{\ell} \lesssim \theta_1.
\end{equation}
Physically, this condition selects waves which propagate at least to the highest 
latitudes $\lambda = \pm (90^\circ - {\theta}_1)$ in the annular domain.
The normalization function
\begin{equation}
N_{\ell}^m = (-1)^m \sqrt{\frac{(\ell-m)!}{(\ell+m)!}\left(\ell+\frac{1}{2}\right)\frac{\pi}{2\Delta\theta}}
\label{eqn:normal}
\end{equation}
where $\Delta\theta \equiv \theta_2 - \theta_1$.
The choice of $\theta_1$ and $\theta_2$ under the constraint of equation (\ref{eqn:orthog}) 
restricts the independence of the
coefficients $A_{\ell m \nu}$ and $B_{\ell m \nu}$ to degrees 
separated by 
\begin{equation}
\label{eqn:deltal}
\Delta{\ell} \approx 2 \pi/\Delta\theta
\end{equation}
\citep{Bogdan1993,Hecht2018}.

The operations needed to estimate the coefficients $A_{\ell m \nu}$ and $B_{\ell m \nu}$ are 
\begin{multline}
\label{eqn:acoeff}
A_{\ell m \nu} = \frac{1}{2 \pi T} \int_{t_1}^{t_2} \int_0^{2\pi} \int_{\theta_1}^{\theta_2} 
W (\theta, \phi) \delta V (\theta ,\phi,t)\\
\times e^{-i(m\phi +2\pi \nu t)} (\Theta_{\ell}^{m})^{*} (\cos \theta) \,d(\cos \theta) \,d\phi \,dt,
\end{multline}
and
\begin{multline}
\label{eqn:bcoeff}
B_{\ell m \nu} = \frac{1}{2 \pi T} \int_{t_1}^{t_2} \int_0^{2\pi} \int_{\theta_1}^{\theta_2} 
W (\theta, \phi) \delta V (\theta ,\phi,t)\\
\times  e^{-i(m\phi +2\pi \nu t)} \Theta_{\ell}^{m} (\cos \theta) \,d(\cos \theta) \,d\phi \,dt,
\end{multline}
where $W$ is the spatial window function and $T = t_2 - t_1$ is the duration of the time interval $(t_1, t_2)$ 
as determined from observational constraints.

For the measurement of meridional circulation, the pole $(\theta = 0)$ is placed at
either the north or south pole of the Sun, and
the resulting distortion between power spectra of poleward 
($|A_{\ell m \nu}|^2)$ and equatorward
($|B_{\ell m \nu}|^2)$ 
propagating waves is typically characterized as
a ``frequency shift''
(at fixed degree and azimuthal order)
of the relevant $f$ or $p$-mode peak.
For short-lived ``local'' modes, with typically high degrees or high frequencies,
and for which peaks in power spectra are blended into continuous
ridges in the $(\ell, \nu)$ domain, this frequency shift can be readily identified as
a shifting of the ridge position. Longer-lived ``global'' modes, typically at low degrees and low frequencies, show isolated peaks, 
whose frequencies are not affected to first order by meridional flow \citep[][]{Gough2010}. 
The ``frequency shifts'' which can be measured in this mode regime through 
FLD methods result from a redistribution of power, through mode coupling produced by
advection, between modes of nearby degree 
and which leak into the observed spectra \citep[][]{Roth2016}.
Our determination of frequency shifts is discussed further in \S\ref{sec.freqshift}.

\subsection{Observations and Initial Analysis}\label{sec.steps}

We use full-disk Dopplergrams, taken with a temporal cadence of once per 45 seconds, obtained from
the HMI instrument onboard the {\it Solar Dynamics Observatory} over an 88 month interval 
spanning 2010 June through 2017 September.
Initial processing of the HMI full-disk Dopplergrams include removal
of the line-of-sight component of solar rotation using a fit to a plane function.
The residual is divided by the cosine of the heliocentric angle to 
account for the (primarily) radially oscillating waves. To avoid foreshortening and other
effects of the extreme limb, we mask out the data beyond a heliocentric angle of 60$^{\circ}$.
A remapping of the line-of-sight Doppler signal to spherical
coordinates is performed using bi-linear interpolation. The coordinate grid has
a spacing of $0.3^\circ$ in both latitude and longitude, which is close to the HMI pixel
size at disk center and oversamples the image elsewhere. The range of latitudes 
considered in our analysis is between 20$^{\circ}$ and 60$ ^{\circ}$ in the 
northern hemisphere and -60$^{\circ}$ to -20$^{\circ}$ in the 
southern hemisphere. At the surface, this latitude range comfortably 
straddles the peak of the meridional circulation as directly inferred in the
photosphere \citep[e.g.][]{Hathaway2010}. This latitude range excludes many, but not
all, active regions. No masking of active regions \citep[e.g.][]{Liang2015a} was performed 
and we note there is no consensus on the use of this operation.

\subsection{Center-to-Limb Artifacts and Control Measurements}\label{sec.ctl}

One of the first manifestations of artifacts, or systematic effects in local helioseismology
mistakenly identified as flows, appear to be 
the high-latitude ``counter cells,'' of opposite direction from the poleward
flows inferred at lower latitudes \citep[][]{Haber2002}. These features were demonstrated to
be of spurious origin by \cite{Gonzalez-Hernandez2006c}. More generalized
large-scale artifacts (hereafter ``pseudo flows'') which vary with distance 
from disk center have subsequently been detected 
\citep[e.g.][]{Braun2008,Duvall2009,
Zhao2012b,Greer2013,Kholikov2014b,Chen2017,Chen2018}. 
As noted by \cite{Duvall2009}, light from the limb 
of the Sun will arrive 2.3 seconds after that emitted from disk center.
This has the consequence of decreasing the observed travel time of waves 
propagating from
the limb toward disk center while increasing the times for waves propagating in
the opposite direction. 
This should give rise to an apparent radially symmetric 
pseudo flow directed towards disk center. 
It has been suggested 
\citep{Baldner2012,Zhao2012b} that 
spatial asymmetries between upward and downward
convective motions can cause pseudo flows of either direction (towards disk center or the limb)
and which depend on heliocentric angle and choice of observable (e.g.\ Dopplergram or 
continuum intensity) or instrument.

Pseudo flows have been observed with a variety of helioseismic methods,
including holography \citep{Braun2008}, time-distance helioseismology 
\citep[e.g.][]{Duvall2009,Zhao2012b,Rajaguru2015,Chen2017,Chen2018}, and ring-diagram analysis \citep[e.g.][]{Greer2013}.
From studies such as these, it appears
that pseudo flows exhibit complex variations (including sign changes) with disk position,
instrumentation, and mode properties including both frequency and wavenumber.
\cite{Rajaguru2020} suggest that the inference of the deep meridional circulation
using time-distance analyses may vary significantly with the choice of temporal frequency of the 
pseudo-flow measurements. This may help
to explain discrepancies in inferences obtained from different groups using the same datasets.

A starting point to correct the data is the assumption that
the pseudo flows are symmetric about the Sun-observer axis. 
For the case of travel-time differences inferred using time-distance methods,
a method first suggested by \cite{Duvall2009}, and subsequently carried out by others
\citep[e.g.][]{Zhao2012b,Kholikov2014b} involves subtracting the travel-time 
differences, measured along the axis connecting the east and west limbs, 
from the measured north-south differences along the central meridian. 

\begin{figure*}
\plottwo{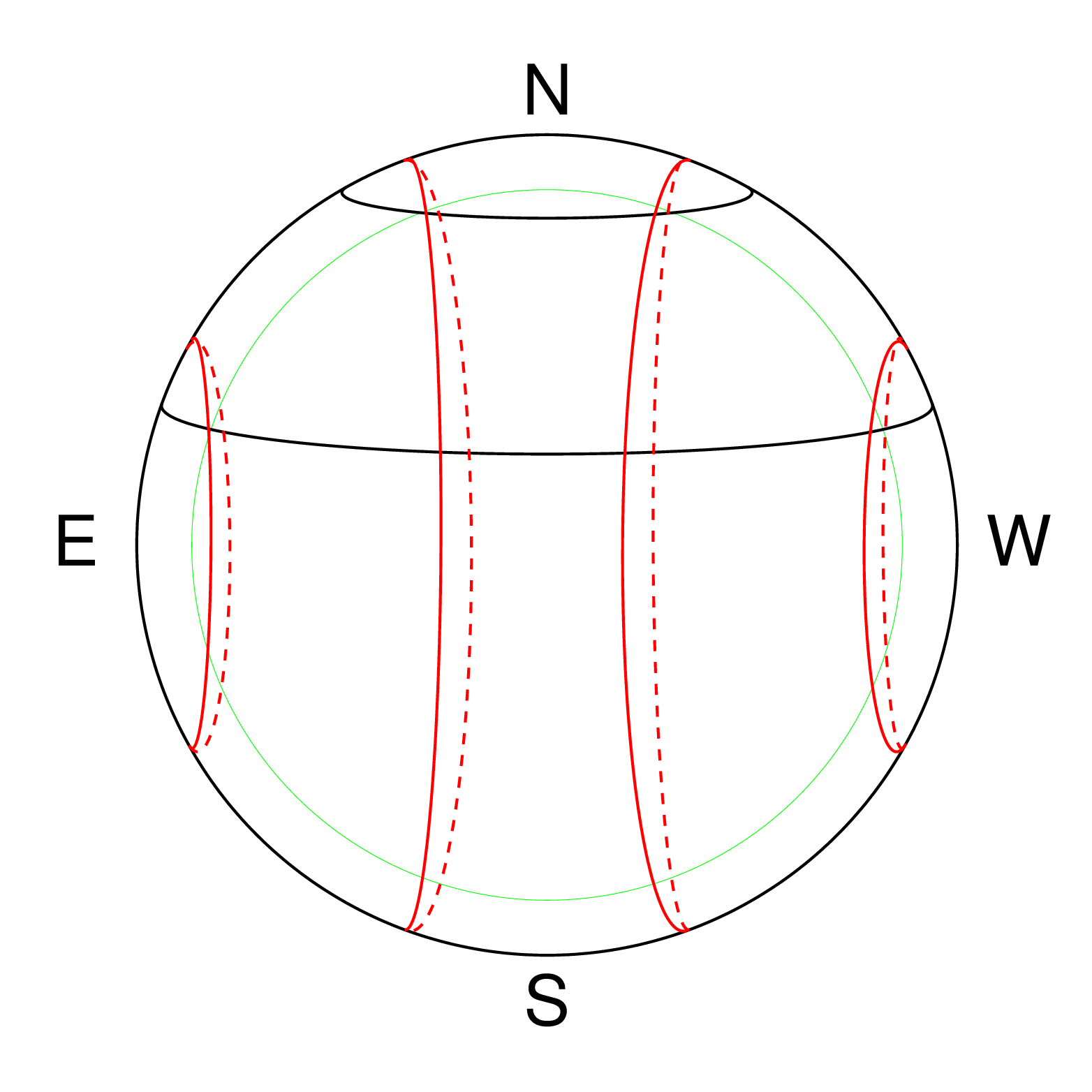}{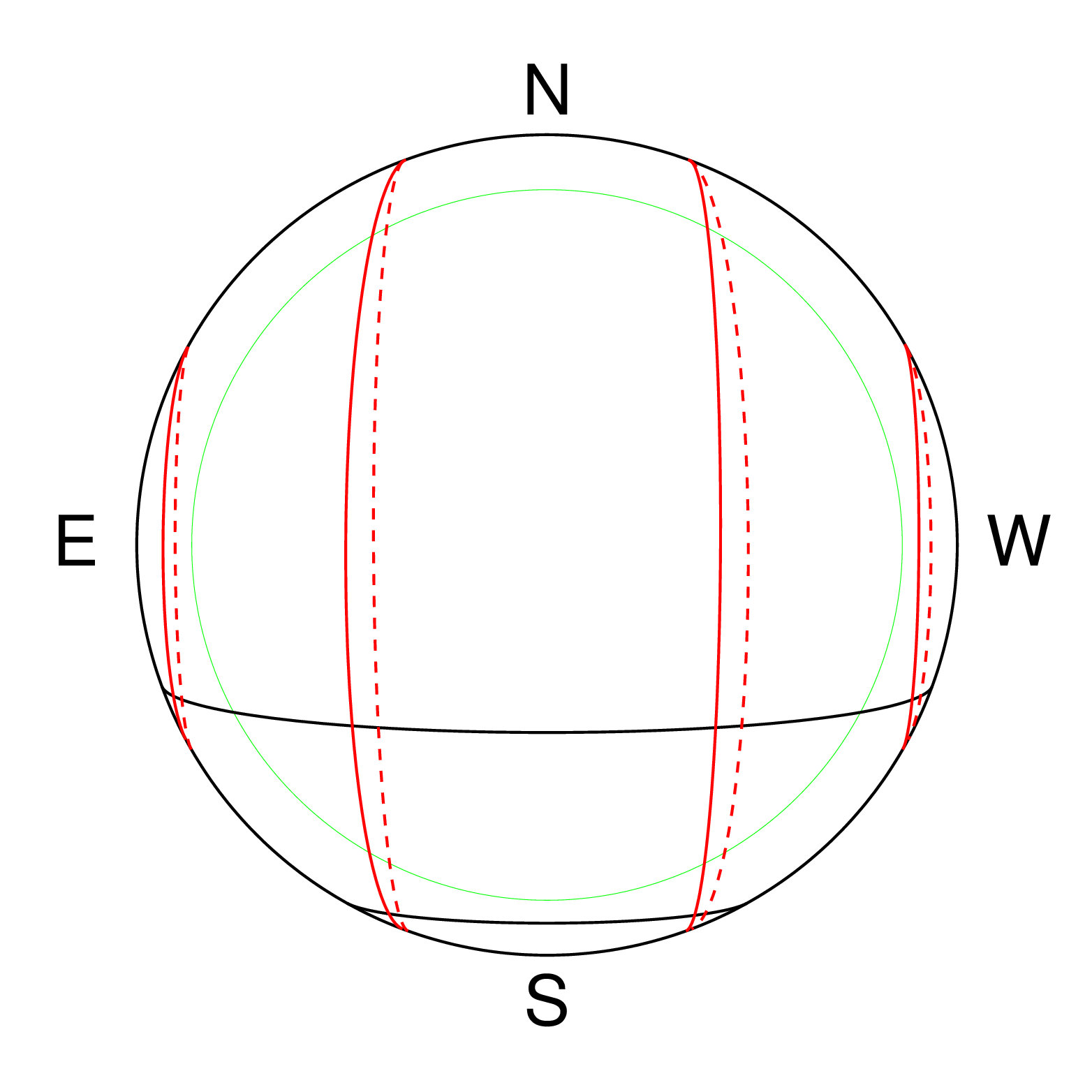}
\caption{
The regions of the Sun for which nominal and control Fourier-Legendre decomposition measurements 
are performed. For this figure, the north pole of the Sun is tilted toward the observer.
The solid black lines in the left (right) panel shows the nominal latitude ranges 
employed in the northern (southern) hemisphere. The boundaries indicate latitudes of
$+20^\circ$ to $+60^\circ$ in the north (left panel) and $-60^\circ$ to $-20^\circ$ in 
the south (right panel). 
The red lines in both panels indicate the regions
analyzed in the control measurements, used to correct the nominal measurements for center-to-limb
artifacts. The solid (dashed) red lines indicate the start (end) of an 8-hour interval
over which the annular control regions are tracked with a Carrington rotation rate.
The green circle indicates a heliocentric angle of 60$^\circ$, beyond which Dopplergram pixels are
excluded from the analysis.
}
\label{fig.geom}
\end{figure*}

In this work, we apply a analogous procedure suitable for the FLD method.
Specifically, a set of control analyses are performed with the geometry rotated 
90$^\circ$ from the true north (or south) poles to analogous locations at 
the east and west limbs (Figure~\ref{fig.geom}). So that the annular regions covered
by these control measurements sample a similar, but rotated, spatial distribution 
of pixels at all times, the positions of the poles of these analyses (hereafter referred to as 
``pseudo-poles'') should mirror the yearly oscillation of the true
poles towards or away from the observer. As discussed further in \S\ref{sec.ctl_shifts}, control analyses 
at both east and west limbs are necessary to separate the effects of the pseudo flow 
from those due to solar rotation.
Furthermore, since we are interested in analyzing the meridional flows in both hemispheres, 
we require a pair of control analyses specifically designed for each hemisphere, 
as illustrated in Figure~\ref{fig.geom}.
If the positions of the pseudo poles were not tracked to follow solar rotation, 
the solar photosphere would rotate substantially across the annular regions in the control 
analyses over any reasonable analysis time frame. The rotational signal introduced would be
up to two orders of magnitude greater than the desired pseudo-flow signals
and the resulting distortion of the power spectra would be detrimental to the analysis. 
Rotating the positions of the pseudo poles at a mean Carrington rate alleviates most of
this distortion, while leaving a smaller effect due to deviations of solar differential rotation
from the Carrington rate.
The cost of this procedure is the requirement to limit the duration of the analyses such
that the geometry does not substantially change over the chosen time span. 
All of these considerations suggest a strategy whereby control measurements are made over 
short intervals while the position of the pseudo poles is tracked at the Carrington rate and
reset back to its initial position on the solar disk at the start of each interval.
Figure~\ref{fig.geom} illustrates the change in the annular regions over 8 hours, which is
the interval chosen for this analysis. 

We use the term ``nominal'' to indicate the measurements used to infer the
meridional flow in the northern and southern hemisphere and distinguish them 
from the ``control'' measurements described above.
For the nominal measurements, which use coordinates aligned with the north and south poles,
we remap from Dopplergram coordinates to Carrington coordinates, thereby applying 
the same tracking. Unlike the control measurements we are free to consider longer time
intervals for the nominal measurements without unwelcome distortions of the power 
spectra due to rotation. For this study, we consider power spectra constructed over
time intervals of 8 hours (as in the control measurements) as well as over
one month intervals. This allows an assessment of potential issues arising from the 8-hr limitation 
of the control measurements. For the ``one-month'' power spectra, we extract and pad data
from each calendar-month with zeros as necessary to compute spectra over a fixed duration of 31.0 days.
The nominal measurements are obtained over the full 88 month duration, while
the control measurements were carried out using the first 24 months of the time interval.
Prior assessments of the pseudo flow \cite[][see their Figure 4]{Liang2018} 
suggest that this systematic effect changes little with time.

The symmetry between the east and west control measurements allow the contribution of
solar differential rotation to be assessed and removed from the center-to-limb aligned pseudo flow 
since the former has opposite effects between the two measurements while the latter has the
same sense in both measurements.
This is analogous to the pseudo-flow assessment used in some time-distance analyses designed
to extract the antisymmetric component of the east-to-west travel-time shifts
within a narrow equatorial strip \citep{Zhao2013}. 
Our control measurements are by design not confined to the equatorial regions, but instead
require an identical, but rotated, annulus which includes 
relatively high-latitude contributions (see, for example, the control annuli in red shown in
Figure~\ref{fig.geom}).
Unfortunately, this results in a small but significant component of the true meridional 
flow leaking (equally) into both control measurements in a fashion which is non-trivial to
separate from the pseudo flows.  To see this, we use the terms ``outward'' 
and ``inward'' to describe the directional sensitivity of the control measurements in 
analogy to the ``equatorward'' and ``poleward'' directions in the nominal measurements.
The leakage of meridional flows into the control measurements results from the fact that,
away from disk center, meridional lines connecting the nominal north
and south poles are not perpendicular to the meridional lines connecting the pseudo poles
at the east and west limbs. Consequently, a fraction of the poleward directed flow in either hemisphere
leaks into the control annuli as an outward directed flow with respect to either (east or west) 
pseudo pole. Alternately, an equatorward flow from either (true) pole 
produces an inward directed leakage in the control measurements.
We estimate the magnitude of this leakage (which is on the order of 20\%) and discuss its mitigation 
in \S\ref{sec.nom_shifts}.

\subsection{Power Spectra}\label{sec.powspec}

The coefficients $A_{\ell m \nu}$ and $B_{\ell m \nu}$ are computed from 
equations (\ref{eqn:acoeff}) and (\ref{eqn:bcoeff}) using numerical integration over
azimuthal angle $\phi$ and colatitude $\theta$ and the use of fast Fourier transforms (specifically,
the FFTW library) in time. The window function $W$ is determined from the latitude ranges
and 60$^\circ$ limb cutoff illustrated in Figure~\ref{fig.geom} within which the data is
apodized with a Welch window in $\theta$ and a raised cosine bell, with a width of 12$^\circ$, in $\phi$.
We compute coefficients for degrees $9 \leq \ell \leq 999$ and
azimuthal orders $-25 \leq m \leq 25$, except for modes with degree
$\ell \leq 45$ for which the highest order is determined from equation (\ref{eqn:impact}). 
The contribution from low frequencies is reduced 
by taking
successive differences in time of the relevant integrand for each $\ell$  and $m$.
Potentially compromised Dopplergrams (e.g. due to cosmic rays or other issues) are identified 
from an examination of the variation in time (over each 8-hour or 31-day interval) of
the sum over $\ell$  and $m$ of the squared values. Anomalous values of this parameter,
defined as less than one-half of, or
greater than twice, the median over the interval, identify problematic Dopplergrams which are then 
removed from the time series through substitution with zeros. Over the 88-month interval, only about
6.6\% of HMI Dopplergrams were either rejected through this criteria or otherwise missing from 
the time series.

\begin{figure*}
\plottwo{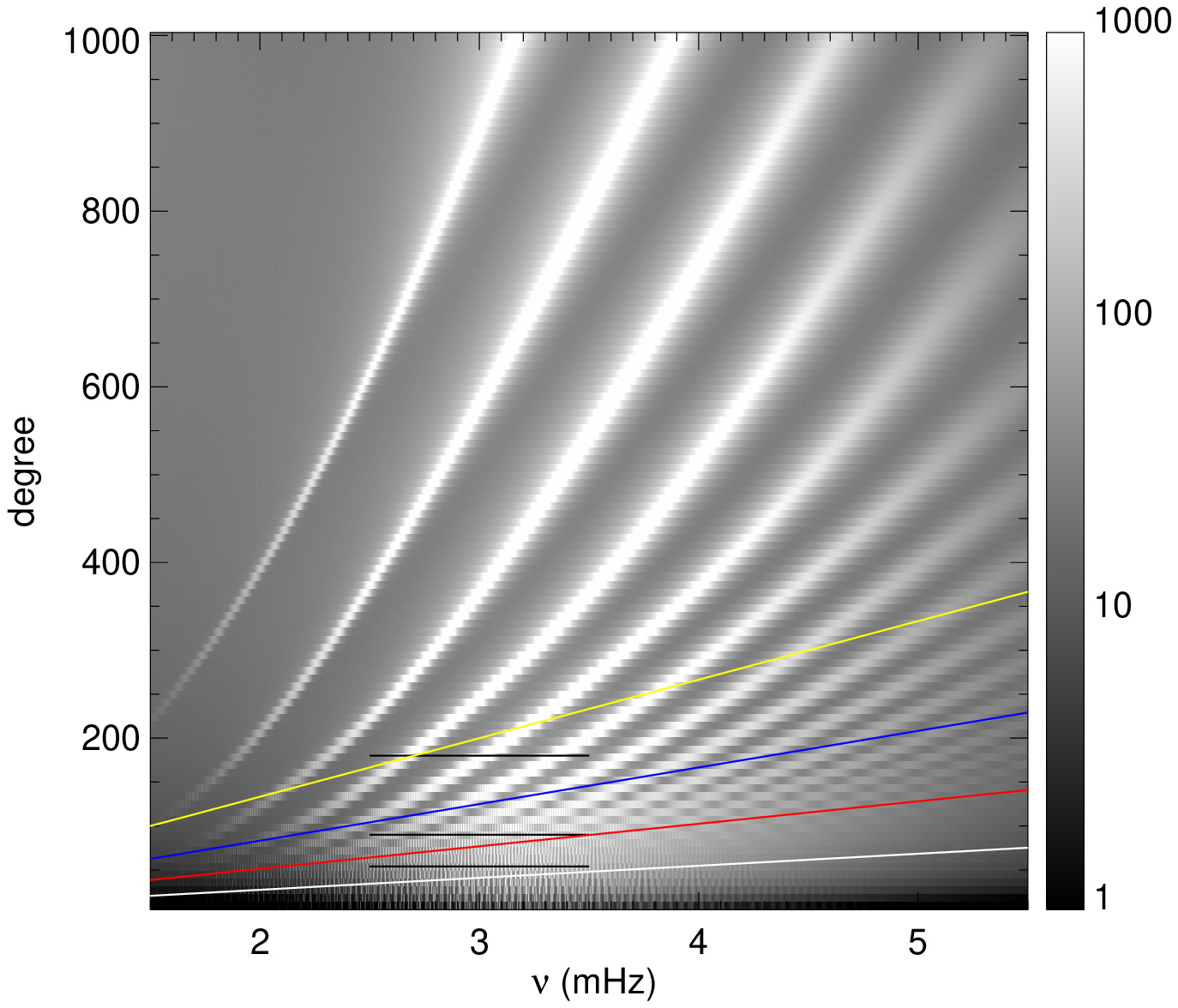}{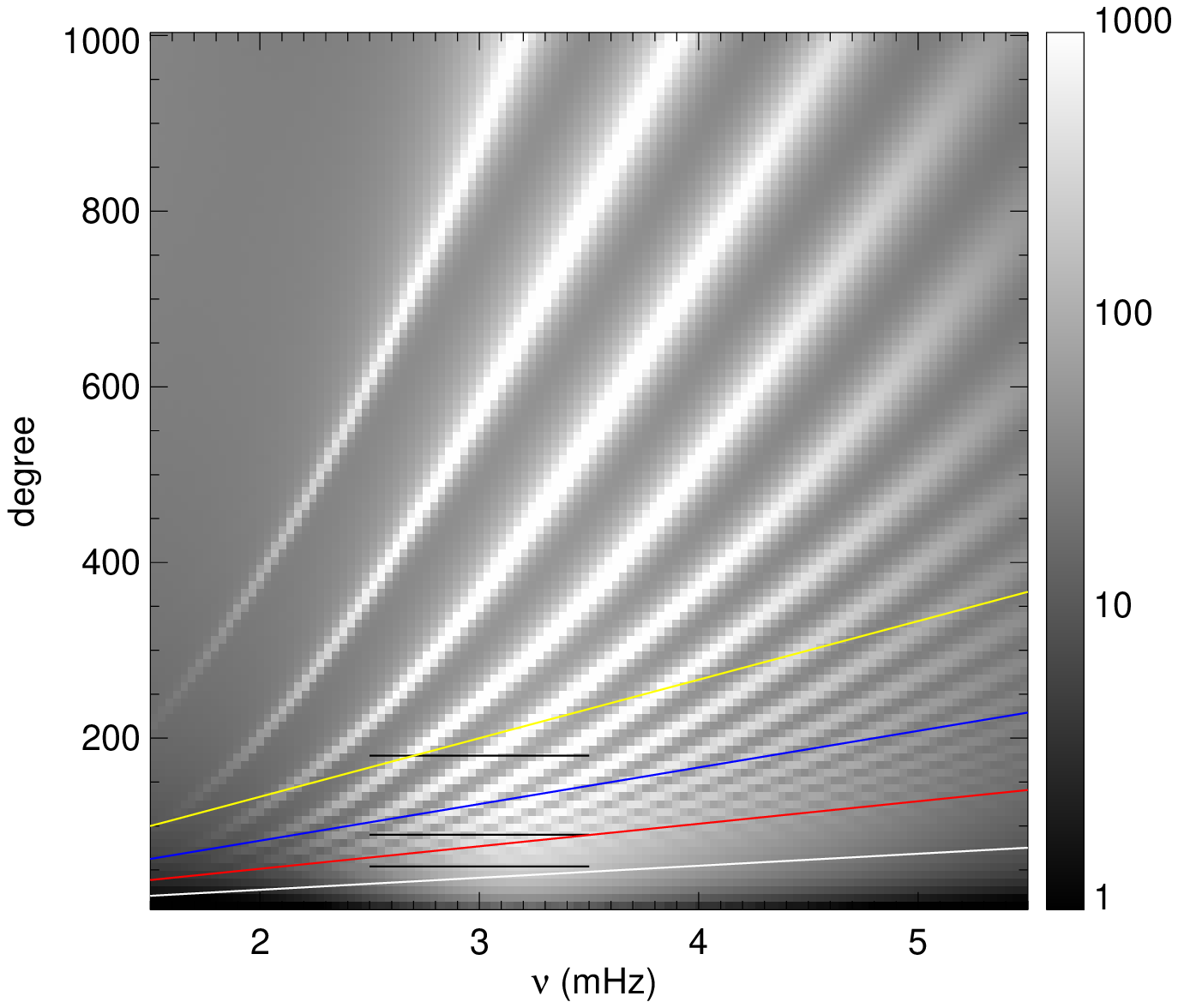}
\caption{
Power spectra of poleward traveling waves in the northern hemisphere computed using Fourier transforms
over 8-hr (left panel) and 31-day (right panel) intervals. and summed over a total duration of 88 months.
Spectra are summed over azimuthal order as described in the text and
averaged over the 88-month duration. The grey scale shows the 
power in arbitrary units in a logarithmic scale. Lines of constant phase speed ($\nu/\ell$) are overlaid,
corresponding to wave lower-turning depths of 25 Mm (yellow), 50 Mm (blue), 100 Mm (red) and 200 Mm (white).
Horizontal black lines indicate slices which are shown in Figure~\ref{fig.pslices}.
}
\label{fig.powspec}
\end{figure*}

\begin{figure}
\plotone{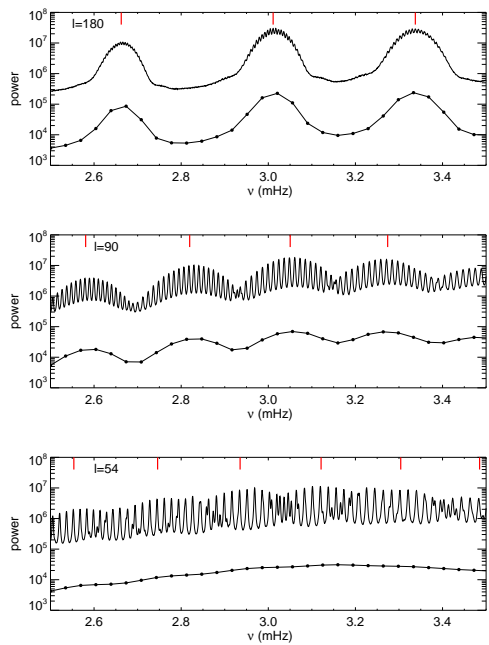}
\caption{
Slices extracted from the power spectra shown in Figure~\ref{fig.powspec} for several values
of $\ell$ as indicated in each panel. The lines connected with filled circles indicate the 
power (in a logarithmic scale) observed in the 8-hr spectra, while lines without dots indicate
power from the 31-day spectra. Red fiducial marks indicate tabulated mode frequencies for different radial orders.
The units are arbitrary and the curves have been displaced vertically
from each other for clarity.
}
\label{fig.pslices}
\end{figure}

Power spectra for poleward ($|A_{\ell m \nu}|^2$) and equatorward ($|B_{\ell m \nu}|^2$)
traveling waves are summed over azimuthal order:
\begin{align}
P_{A}(\ell, \nu) \equiv \displaystyle\sum_{m=-\mu}^{\mu} |A_{\ell m \nu}|^2; \\
P_{B}(\ell, \nu) \equiv \displaystyle\sum_{m=-\mu}^{\mu} |B_{\ell m \nu}|^2, 
\label{eqn:powersum}
\end{align}
where $\mu = \min[25, \mathrm{floor}(\theta_1\ell)]$ and the floor function is the greatest integer
less than or equal its argument.  This summation is justified since 
the sensitivity of the frequency shifts is very nearly independent of $m$, as discussed further in
\S\ref{sec.forward}. 
Spectra for each 8-hr or 31-day interval are then summed over time. The nominal measurements are
summed over the entire 88-month duration and also over one-year intervals for each calendar year 
2011 through 2016. The latter are used to 
estimate errors in the frequency shift measurements (\S\ref{sec.freqshift}), and
to examine the temporal variation of the shifts (\S\ref{sec.year2year}).
Figure~\ref{fig.powspec} shows power spectra for poleward traveling waves over the northern hemisphere
and averaged over the 88-month duration, as determined over 8-hr and 31-day intervals.  
It is difficult to ascertain differences between the two spectra when displayed as images with similar
scales and dimensions, so line plots of the spectra of selected degrees are directly compared in
Figure~\ref{fig.pslices}.

To extract the frequency shifts for inferring the meridional circulation we have chosen to
use the 8-hr power spectra. This ensures that the shifts are determined uniformly for both the
nominal and control measurements, the latter requiring the primary limitation on the temporal duration
as described earlier. The comparison of the spectra between 8-hr and 31-day intervals is
instructive, however. For modes with degree greater than about 150, there is little difference
in the appearance of the mode ridges other than the difference in the frequency sampling. This
can be seen in the top panel of Figure~\ref{fig.pslices} for example. As one examines the spectra
at successively lower $\ell$, it becomes apparent that global peaks are resolved in the 31-day spectra but
not in the 8-hr spectra. These multiple narrow peaks represent modes with nearby $\ell$ which leak
into the power spectra due to the window function which isolates a 40$^{\circ}$ strip in latitude). 
Distinct, if blended, ridges corresponding to different radial orders
are visible in the 8-hr spectra above the red line in Figure~\ref{fig.powspec}, 
which corresponds to constant
phase speed $\nu/\ell = 0.039$ mHz. This phase speed corresponds to a lower turning point depth of
100 Mm, or about midway in the convection zone. The middle panel of Figure~\ref{fig.pslices}
shows an example of the power spectra in this regime. Below the red line in Figure~\ref{fig.powspec}
it is clear that distinct ridges for each order are not resolved (e.g.\ bottom panel of 
Figure~\ref{fig.pslices}). This is due to the overlap of
the leaked modes between neighboring radial orders which is more readily observed in the 31-day spectra.

\subsection{Multi-Ridge Fitting}\label{sec.mrf}

To extract the frequency shifts between poleward and equatorward traveling waves from the power spectra
(as well as the inward and outward traveling waves in the control spectra) we compared three different
methods. These included 1) determining the difference between the centroid frequencies of poleward
and equatorward ridges \citep[e.g.\ as carried out by][]{Braun1998b}, 2) finding the peak in the
cross-correlation function between the two ridges, and 3) fitting the power spectra to models which
account for the shape of the ridges and their shifts in frequency.
Comparison of the different methods as well as tests with artificial power spectra are described
in Appendix~\ref{sec.append1}. These tests demonstrated the presence of systematic biases in the first
two methods above which are due to the contamination from ridges with neighboring radial orders. 
Methods which simultaneously fit multiple ridges can account for the blending of nearby ridges 
in a manner not possible with those which fit isolated ridges \citep[e.g.][]{Greer2014}.

\begin{figure}
\plotone{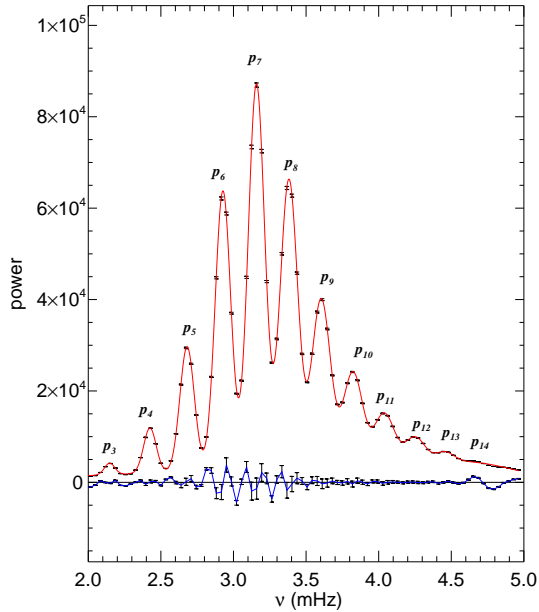}
\caption{
A sample fit to a slice (at $\ell = 99$) of the 88-month averaged (8-hr interval) power spectra 
shown in the left panel Figure~\ref{fig.powspec}. The observed power spectra are shown as black
error bars, with the error determined from year-to-year fluctuations in the spectra. The red curve
shows the resulting fit. 
The $p$-mode ridges are labeled as ``$p_n$'' where $n$ is the radial order.}  
The blue curve shows the residuals (observations minus fit) with the same
error bars as observed in the power spectra. The residuals and its error bars have been multiplied by
a factor of 5 for clarity.

\label{fig.samplefit}
\end{figure}

To carry out multi-ridge fitting (hereafter ``MRF''), we first extract the power at each $\ell$
as a function of frequency. Using lookup tables of mode frequencies we identify the number of ridges, $N$,
with differing radial order $n$ present below 5 mHz. 
We then fit the power to a sum of $N$ Gaussians, with free parameters of amplitude, width,
and central frequency, and a background term consisting of a cubic polynomial. The frequency shift of 
interest is 
\begin{equation}
\label{eqn:deltanu}
\Delta\nu(\ell,n) = \nu_A(\ell,n) - \nu_B(\ell,n),
\end{equation}
where $\nu_A$ and $\nu_B$ are the central Gaussian frequencies from the fits to spectra $P_A$ and $P_B$ respectively
for the ridge of degree $\ell$ and radial order $n$.
The fitting code employs MPFIT routines \citep{Markwardt2009} which carry out a non-linear least-squares fit
with uniform weighting. Fits using a weighting by the 
inverse of the square of the error, as determined from year-to-year
fluctuations in the spectra, were also attempted. These 
fits generally underestimated the peak values
and were judged inferior to fits using uniform weighting.  Fits to Lorentzian functions were also attempted,
but these tended to greatly overestimate the troughs between the ridge peaks and forced a negative
background term. This suggests that the Lorentzian functions have line wings which are too large compared to
the observed spectra.
Initial guesses for the Gaussian parameters were
obtained from fitting individual Gaussian functions to isolated ridges for each radial order. 
Constraints on the fits include 1) keeping the Gaussian amplitudes and widths positive and 2) keeping the 
central frequencies within a window spanning the mid-points to adjacent radial orders. 
Figure~\ref{fig.samplefit} shows an example of our modeling result for degree
$\ell = 99$. 
In some instances, ridges at high temporal
frequency could not be fit satisfactorily. An example of this failure is illustrated by the
$p_{14}$ ridge in Figure~\ref{fig.samplefit}.

\section{Frequency Shifts}\label{sec.freqshift}

The frequency shift in equation (\ref{eqn:deltanu}) denotes the difference 
between the central ridge frequency of the poleward 
(or inward) traveling waves minus that of the equatorward (or outward) traveling waves.
It is convenient (particularly in regards to the modeling efforts described in \S\ref{sec.forward})
to define a scaled frequency shift $U$:
\begin{equation}
\label{eqn:udefine}
U(\ell,n) \equiv \pi R_\odot \Delta\nu(\ell,n)/\ell
\end{equation}
where $R_\odot$ is the solar radius. The units of $U$ are that of speed (e.g.\ m s$^{-1}$) and
$U/R_\odot$ has a physical meaning as the equivalent uniform (i.e.\ ``solid-body'') 
angular rotational speed of a shell of a star needed to yield a given frequency shift $\Delta\nu$
of waves fully confined within the shell. In the application here, it is useful to think of
this hypothetical shell as roughly spanning the solar surface to the depth of the lower turning point of a given
mode $(\ell,n)$. Inferring the actual radius (depth) dependence of the flow from the collection of measurements
$U(\ell,n)$ is discussed in \S\ref{sec.forward} but, at least for modes with shallow turning points (e.g.\ large $\ell$), 
$U$ provides an estimate of the local flow speed as experienced by those waves.
The sign of the frequency shift is such that a positive value of $U$ indicates a flow (or pseudo flow)
which is directed poleward (or inward).

Errors in the determination of $U$ were obtained by dividing the 88-month dataset
into yearly intervals and defining the errors to be the year-to-year root-mean-squared fluctuations in the
frequency shifts divided by the square-root of the total time interval in years.
In general, the errors at phase speeds below about 0.015 mHz are of order 1 m s$^{-1}$, but increase 
sharply with increasing phase speed to nearly 10 m s$^{-1}$. 
A common mode set, consisting of nominal measurements
with an error less than 10 m s$^{-1}$ for both hemispheres, was established. This mode set spanned
a range in degree $81 \leq \ell \leq 999$, frequencies between 2.0 and 5.0 mHz, and radial orders
from $n=0$ to 12. Among the 565 modes in the final dataset, the deepest penetrating modes
have phase speeds around 0.04 mHz and lower turning point depths around 100 Mm below the surface.

\subsection{Control Measurements}\label{sec.ctl_shifts}

\begin{figure}
\plotone{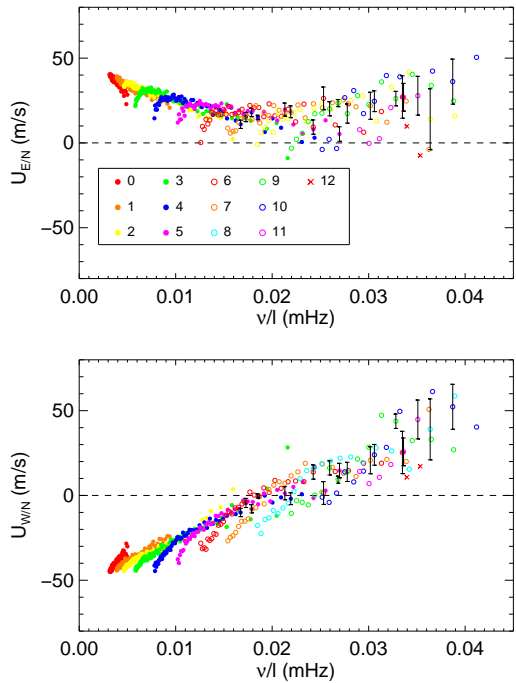}
\caption{
The scaled frequency shifts for the east and west-limb control analyses as designed to
correct the northern hemisphere, as a function of
the phase speed $\nu/\ell$. The top (bottom) panel shows the results $U_{E/N}$ ($U_{W/N}$) 
for the west (east) control measurements respectively.
Different values of radial order $n$ are designated with colors
and symbols as indicated. Error bars are shown for a selected sample of modes, and only for
phase speeds above 0.015 mHz, for clarity. Errors for measurements at lower phase speeds
are on the order of the symbol size.
}
\label{fig.uew}
\end{figure}

Figure~\ref{fig.uew} shows the scaled frequency shifts for the east and west control
analyses which were designed to mirror the (rotated) position of the latitude annulus to
correct the northern-hemisphere measurements. We refer to these east and west control measurements 
as $U_{E/N}$ and $U_{W/N}$ respectively. Results for the control analyses designed for the
southern hemisphere (not shown) are similar. For phase speeds below 0.02 mHz,
$U_{E/N}$ and $U_{W/N}$ have opposite sign and exhibit linear trends with
phase shift which have slopes of opposite sign.  The sign of these shifts is indicative of
a net eastward directed flow, which is retrograde with respect to solar rotation,
and can be identified as the residual effects of differential
rotation relative to the tracked Carrington rate.
At phase speeds above 0.02 mHz, the results
for both control measurements show positive shifts approaching nearly 50 m s$^{-1}$.

\begin{figure}
\plotone{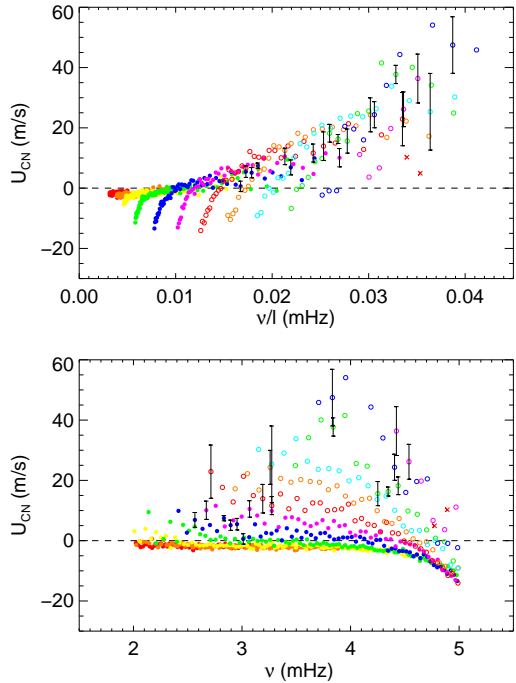}
\caption{
The scaled frequency shift $U_{CN}$ for the combined (east and west) control analyses
designed for the northern hemisphere. The top panel shows the results as a function of
phase speed while the bottom panel shows the results as a function of mode frequency.
Colors and symbols show different radial orders
as indicated in Figure~\ref{fig.uew}.
}
\label{fig.ucn}
\end{figure}

Adding the frequency shifts for the east and west control measurements 
cancels out the effects of rotation such that the remaining
shifts are primarily caused by the center-to-limb effect. 
However, an alternative way of achieving this is to sum the relevant inward or outward
power spectra over both control measurements and extracting the frequency shifts, $U_{CN}$,
from the resulting combined spectra. This method results in incrementally less noise than
frequency-shift subtraction and was employed here.
Figure~\ref{fig.ucn} shows $U_{CN}$ plotted as functions of phase speed (upper panel)
and mode frequency (lower panel).
Results for the two control measurements designed for the southern hemisphere (not shown) are similar.
The pseudo flow revealed by these measurements contains both outward and inward directed 
components.  As the phase speed of the modes increase, the pseudo flow increases from 
inward directed with amplitudes around -2 m s$^{-1}$ to outward directed 
with amplitudes reaching up to 50 m s$^{-1}$ for the deepest modes. 
Viewed as a function of frequency, each radial order shows pseudo flows
which decrease and become inflows with increasing frequency.
Many of the radial orders converge to a common function of frequency above 4.5 mHz. 
A remarkably similar result for the frequency and phase-speed variation of
the center-to-limb effect was found by \cite{Greer2013} using ring-diagram
analysis of HMI data with 30$^\circ$ tiles. The frequency variation of
the center-to-limb effect has also been explored using a Fourier analysis of time-distance
correlations \citep{Chen2018}. Our results are at least qualitatively similar to
these results which include a change of sign of the pseudo flow near 5 mHz.

\subsection{Corrected Frequency Shifts}\label{sec.nom_shifts}

\begin{figure}
\plotone{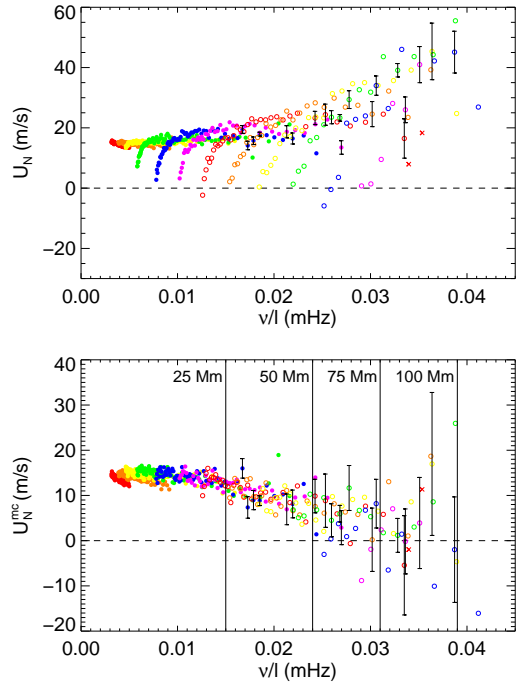}
\caption{
The nominal frequency-shift measurements for the northern hemisphere. The top panel shows
the raw frequency shifts without correction for the center-to-limb effect. The bottom
panel shows the results after a subtraction of the frequency shifts as determined from
the east and west control measurements and multiplication by a constant which takes into
account the leakage of the meridional-flow signal into the control measurements (see text).
Vertical lines show the phase speeds corresponding to the mode lower-turning depth  
The colors and symbols
show different radial orders as indicated in Figure~\ref{fig.uew}.
}
\label{fig.unom}
\end{figure}

Figure~\ref{fig.unom} shows the nominal frequency-shift measurements for the northern
hemisphere before and after a correction for the center-to-limb effect. The shallowest
modes show raw frequency shifts of around 15 m s$^{-1}$, which can be identified
with the meridional flow. However, the entirety of the measurements resemble those
obtained from the control measurements (the top panel of Figure~\ref{fig.ucn}) but offset vertically.
Subtracting the latter from the raw measurements removes most of the center-to-limb
effect, but this operation alone fails to account for the leakage of the true meridional
flow signal into the control measurements as discussed at the end of \S\ref{sec.ctl}.

To appropriately include this effect in our correction, we assume the
control measurements $U_{CN}$ and $U_{CS}$ contain a component due to leakage which is
proportional by a factor $f_{\rm lk}$ to the signal produced by the 
meridional circulation in the nominal 
measurements. As discussed in Appendix~\ref{sec.append2} this suggests
a modified operation to retrieve the desired shifts due to
meridional circulation, $U_{N}^{\rm mc}$, from the raw shifts $U_N$:
\begin{equation}
\label{eqn:modcorr}
U_{N}^{\rm mc} =  (1-f_{\rm lk})^{-1} (U_N - U_{CN}).
\end{equation}
We assume $f_{\rm lk} = -0.19$ which is estimated from an evaluation of 
surface measurements of the meridional circulation and the assumption that the 
leakage does not vary with phase speed among the mode set. This estimate is discussed and 
justified in Appendix~\ref{sec.append2}. Consequences
of a deviation of $f_{\rm lk}$ from this constant value on our derived meridional flow profiles
are discussed in \S\ref{sec.discussion}. The bottom panel of Figure~\ref{fig.unom} 
shows the results of our
correction for the northern hemisphere (similar results for the southern hemisphere are not
shown) and represent the frequency shifts we consider in
our forward modeling efforts.

\section{Forward Modeling}\label{sec.forward}

\begin{figure}
\plotone{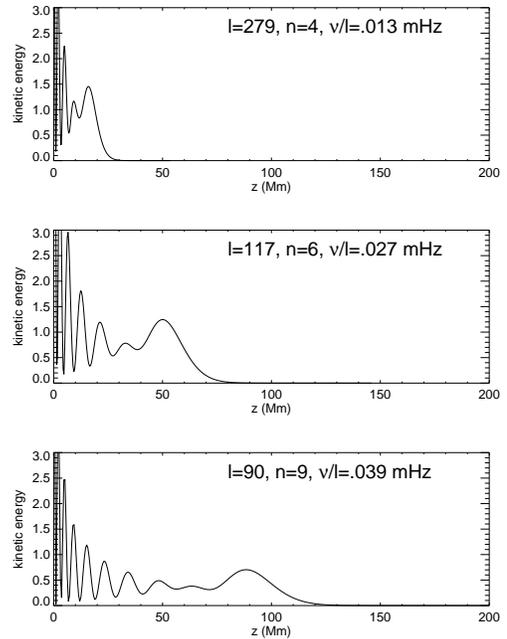}
\caption{
Some examples of the depth variation of the kernels, which are proportional to 
mode kinetic energy density.
The depth $z$ is defined as $z = R_\odot - r$.
From top to bottom, the panels show the 
kernels for successively deeper penetrating modes, with mode properties
as indicated in each panel.
}
\label{fig.kernels}
\end{figure}

The goal of the modeling is to infer the depth variation of the latitude-averaged
meridional flow
\begin{equation}
\label{eqn:vthet}
\left\langle v_\theta \right\rangle (r) \equiv \frac{1}{\Delta\theta} \displaystyle\int_{\theta_1}^{\theta_2} 
   v_{\theta}(\theta, r) d\theta 
\end{equation}
in each hemisphere. The flow is related to the frequency shifts $U^{\rm mc}$, with the subscript $N$ or $S$ omitted but understood, as
\begin{equation}
\label{eqn:kernel}
U^{\rm mc} (\ell, n) = \left\langle\frac{S(\ell,m)}{S(0,m)}\right\rangle_{m} 
         \frac{ {R_\odot} \displaystyle\int_0^{R_\odot} \frac{\left\langle v_\theta \right\rangle}{r} K_{\ell n}(r) dr}{\displaystyle\int_0^{R_\odot} K_{\ell n}(r) dr},
\end{equation}
where the kernels $K_{\ell n}(r)$ represent the kinetic energy density 
\citep{Gough1983,Birch2007b} and have been used in prior FLD modeling \citep{Braun1998b,Roth2016}.
The function $S(\ell,m)$ describes the sensitivity of the shifts with azimuthal order 
(see Appendix~\ref{sec.append3}) while the
angular brackets around the ratio $S(\ell,m)/S(0,m)$ represent an average over the same azimuthal orders $m$ 
used to sum the power spectra from which the scaled shifts were extracted. 
Figure~\ref{fig.kernels} shows some examples of the kernels $K_{\ell n}(r)$.

\begin{figure}
\plotone{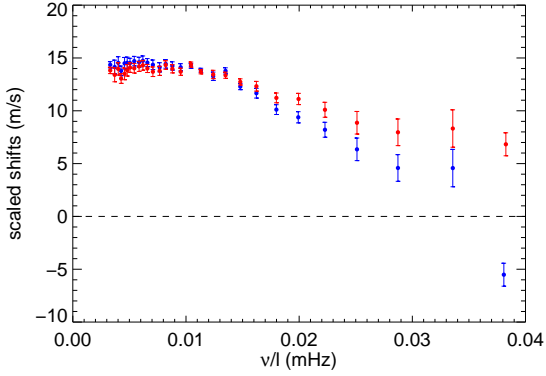}
\caption{
Averages of the corrected scaled frequency shifts
computed within narrow bins in phase speed, with error bars defined by the standard
error of the mean. The blue (red) points with errors show the results for the northern (southern)
hemisphere respectively. 
}
\label{fig.binned_shifts}
\end{figure}

We employ a ``forward modeling'' approach whereby the observed frequency shifts 
are directly compared with results obtained from equation (\ref{eqn:kernel}) using 
different choices of the flow $\langle v_\theta \rangle$.
We first average the measurements in narrow bins in phase speed to obtain the mean 
and its standard error. The bins are chosen to isolate groups of twenty
frequency-shift measurements. 
Figure~\ref{fig.binned_shifts} shows the averaged values, with error bars given
by the standard error of the mean.
The averaging reveals systematic differences between the two hemispheres. At phase speeds
below 0.01 mHz, the corrected shifts in the northern hemisphere exceed those in the south
by an amount just under 1 m s$^{-1}$. At a phase speed near 0.01 mHz the shifts for both
hemispheres are in agreement, but at higher phase speeds the shifts in the northern hemisphere
are significantly smaller than in the south with the difference increasing with phase speed.
At a phase speed near 0.03 mHz, the results differ by about 4 m s$^{-1}$ with the
results in the south being roughly twice that in the north. The biggest difference
between hemispheres occurs at the highest phase-speed bin, where the northern shift is
of opposite sign than those in the south.

\begin{figure}
\plotone{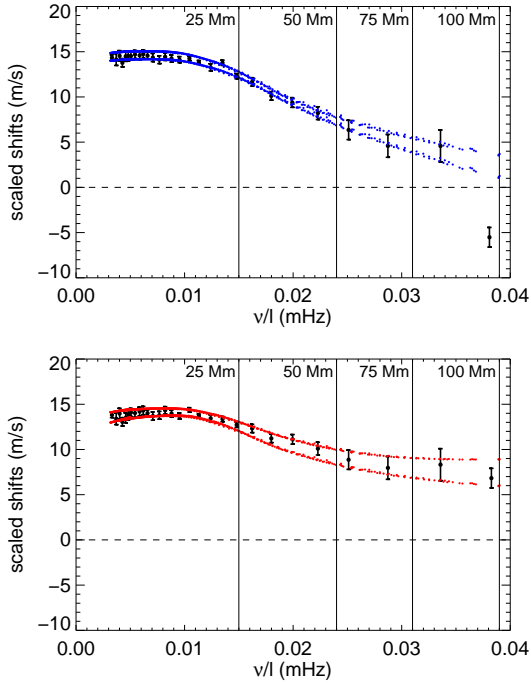}
\caption{
Black filled circles represent averages of the corrected scaled frequency shifts
computed within narrow bins in phase speed, with error bars defined by the standard
error of the mean. The top (bottom) panel show the results for the northern (southern)
hemisphere respectively. In each panel, the colored dots show the predicted scaled shifts from 
a pair of flows selected to straddle the range of the observations 
caused by the standard errors. 
Vertical lines indicate different lower turning point depths as indicated.
}
\label{fig.forward}
\end{figure}

For each hemisphere, two flow profiles were obtained by trial and error. The goal
is that the two flow functions represent plausible fits to the 
upper and lower limits of the mean shifts as determined by their standard errors.
In general, the predicted scaled shifts corresponding to the 565 observed modes 
for a given flow lie with very little scatter 
around a relatively smooth and continuous function of phase speed (Figure~\ref{fig.forward}).
This facilitates a rapid assessment of a candidate flow through 
visual inspection of plots such as Figure~\ref{fig.forward}. 

\begin{figure}
\plotone{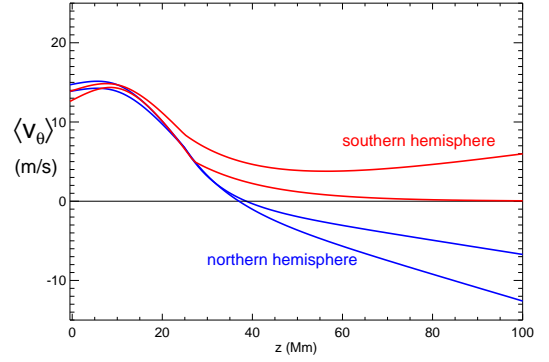}
\caption{
The final four meridional flow functions whose predicted scaled frequency shifts are
shown in Figure~\ref{fig.forward}. 
The depth $z$ is defined as $z = R_\odot - r$.
The two red (blue) curves show plausible limits of
the flows in the southern (northern) hemisphere roughly consistent with the errors
of the observations.
}
\label{fig.flows}
\end{figure}

We use a characterization of the flow which consists of a cubic spline interpolation
between three prescribed anchor points spanning depths between 0.5 Mm above the photosphere
and approximately 25 Mm below the photosphere (this lower depth varied somewhat among
the four fits). The speed at the three anchor points was determined through trial and error
and involved fixing the shallowest points first and adjusting successively deeper 
values until suitable matches between observed and predicted frequency shifts were obtained.
Below the deepest anchor point  
the flow function was assumed for simplicity to contain of a sum of exponential and linear 
terms. The flow speed at 
the deepest anchor point fixes the amplitude of the exponential term while the linear
term is set to zero at this depth. There are consequently two free parameters (the decay scale of
the exponential and the slope of the linear term) which determine the depth dependence of
the flow below this depth.
Figure~\ref{fig.flows} shows the final four flows, from which the predicted shifts
shown in Figure~\ref{fig.forward} were obtained. 
For the results in the northern hemisphere, the (negative) bin-averaged shift at the highest phase
speed in Figure~\ref{fig.forward} was ignored, as no reasonable flow model considered could 
reproduce this apparently anomalous measurement.
The model flows shown in Figure~\ref{fig.flows} are similar between
hemispheres over the shallow regime over which the flows were characterized by 
cubic splines. However significant differences are observed below 25 Mm, with the 
most surprising result being the onset of a relatively shallow 
equatorward return flow in the northern
hemisphere at around 40 Mm. Results for the southern hemisphere are consistent with a 
decrease in speed with depth, but remain poleward over the depths considered.

\section{Temporal Variations}\label{sec.year2year}

To further explore the apparent hemispheric differences in the frequency shifts
illustrated in Figure~\ref{fig.binned_shifts}, we compare the variation of this
difference with hemispheric differences in solar activity.
As discussed in \S\ref{sec.powspec}, errors were estimated from frequency shifts 
assessed from power spectra averaged over yearly intervals for each calendar year.
These year-to-year measurements also allow their temporal variations to be determined.
To clearly see general trends while maintaining some discrimination with mode depth, 
the shifts are averaged over mode sets corresponding to three ranges in phase-speed. 
We establish three sets, such that the phase speed $\nu/l$ ranges from 0 to 0.01 mHz
in the ``shallow'' set, 0.01 to 0.02 mHz in the ``intermediate'' set, and 0.02 to 0.04 mHz
in the ``deep'' set. We note the year-to-year frequency shifts are not corrected
for either the center-to-limb effect or the leakage, which does not effect 
the relative temporal changes we are interested in examining.

\begin{figure}
\plotone{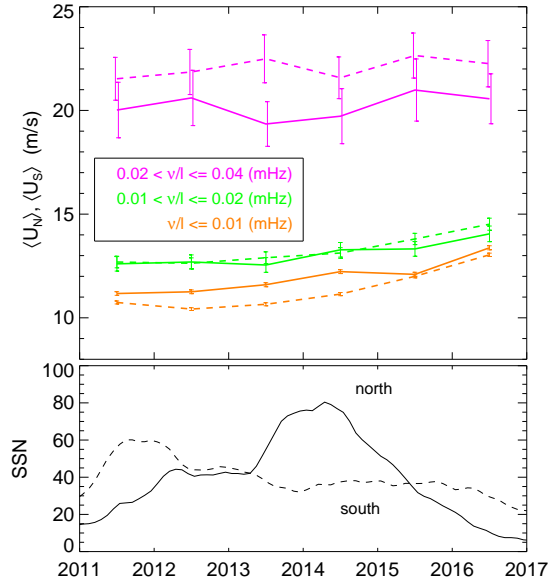}
\caption{
The time variation of nominal, uncorrected, frequency-shift measurements averaged
over different mode sets (top panel), 
and smoothed sunspot numbers for each hemisphere (bottom panel).
The frequency-shift measurements are made over one-year intervals spanning six calendar years
and averaged over phase speed intervals as indicated by the legend. 
Deep, intermediate, and shallow mode-set averages are represented by magenta, green and orange
lines respectively, with solid (dashed) lines indicating results for the northern (southern)
hemisphere. The smoothed sunspot number (SSN) is shown for each hemisphere.
}
\label{fig.year2year}
\end{figure}

Figure~\ref{fig.year2year} shows the results of this averaging, which is done
for both hemispheres, along with the monthly smoothed sunspot numbers 
\citep[hereafter SSN;][]{sidc}. The shallow and intermediate mode-averaged shifts, for
both hemispheres, show gradual increases above the errors over the six years examined,
while the shifts deepest set remain constant within the errors. 
This variation is plausibly related to solar-cycle changes in the meridional
circulation in this cycle \citep[e.g.][]{Zhao2014}. 
Of particular relevance is that the hemispheric
differences which were found in the total 88 month set are present for most, if
not all, of the period shown.
The tendency for the shallow mode shifts in the north to exceed those in the south,
is observed for most of the interval, although this difference nearly disappears
for the last two years. On the other hand,
the north-south discrepancy in the deepest mode shifts is notably present with similar
magnitude (roughly two seconds) during all six years.
During the same interval, solar cycle 24 was initially dominated by a distinct peak of 
sunspot numbers in the northern hemisphere, followed over two years later by a peak in sunspot
number in the south.  This would argue against the hemispheric differences in
the frequency-shift measurements being related in a simple way to asymmetries in the
distribution of magnetic regions.

\section{Discussion}\label{sec.discussion}

Taken at face value the results as illustrated by Figures~\ref{fig.forward} and
\ref{fig.flows} are unexpected in that the shallow return flow present in the north
hemisphere appear to be inconsistent with most prior determinations of the meridional
circulation which employ multi-year observations and which attempt to remove the effects
of center-to-limb related artifacts. 
Specifically, most analyses involving modeling time-distance travel-time perturbations
show a poleward flow in both hemispheres which persist from the surface down to approximately
0.9 $R_\odot$ (i.e. about 70 Mm below the surface). This includes the analyses of
\cite{Zhao2013,Jackiewicz2015,Rajaguru2015,Chen2017}
A notable exception is the recent analysis of \cite{Gizon2020} which is discussed below.
In general, uncertainties in the
inference of the flow increase with the depth of those inferences. This makes the
discrepancy with other helioseismic analyses at a relatively shallow depth all the
more puzzling. If we assume that the results for at least the northern hemisphere
are spurious, than the challenge is to understand the cause of what appears to
be an uncorrected artifact. Of course, real hemispheric differences in the meridional circulation
may very well exist, but it is worth considering the possible nature of pseudo-flows which
are themselves asymmetric.

The comparison of the time variation of the frequency shifts with the sunspot numbers for
both hemispheres (Figure~\ref{fig.year2year}) indicates that the observed hemispheric
differences in the 88-month interval persist regardless of the level
of solar activity or its preference to one hemisphere over the other. While this doesn't
rule out the presence of real solar-cycle variations in flows 
\citep[e.g.][]{Chou2001,Zhao2004,Gonzalez-Hernandez2008,Hathaway2010}, 
or potential pseudo-flow 
artifacts related to magnetic regions \citep[e.g.][]{Liang2015a}, there is no direct
correlation between hemispheric differences in solar activity and the differences in the
frequency shifts.

Errors in the control measurements due to uncertainties of the leakage
discussed in Appendix~\ref{sec.append2} appear to be too small to account for the
hemispheric differences observed in Figure~\ref{fig.binned_shifts}. In particular 
we note that likely errors in the leakage factor $(1-f_{\rm lk})^{-1}$ amounting to 20\% of
the selected value of 0.84 (see Appendix~\ref{sec.append2}) cause uncertainties
in the deepest scaled frequency shifts of around 1 m s$^{-1}$. These are smaller than the
standard errors of the observations shown in Figure~\ref{fig.binned_shifts} and substantially 
smaller than the 4 m s$^{-1}$ discrepancy observed at the highest phase speeds.

It is plausible that the results may be compromised by 
incorrect assumptions regarding the azimuthal invariance of the 
center-to-limb correction. At the deepest
phase speeds, the frequency shifts due to the pseudo flow shown in the top panel
of Figure~\ref{fig.ucn} 
have amplitudes that exceed by about a factor of ten the expected shifts due
to meridional circulation in the nominal measurements (bottom panel of 
Figure~\ref{fig.unom}). Thus, relative differences between the pseudo-flow at either true pole
and the east and west limbs amounting to only 5-10\% could produce 
50-100\% uncertainties in the corrected shifts, and is sufficient to account for
an anomalous hemispheric difference consistent with observations.

Of particular relevance is the recent analysis of \cite{Gizon2020} which compared inferences
of the meridional circulation obtained from three data sources (MDI, GONG, and HMI).
While they generally found the results obtained from MDI and GONG were consistent
with a single circulation cell in both hemispheres, with a return flow present 
at 0.8 $R_\odot$ (approximately 140 Mm below the surface), they also found systematic
errors unique to the measurements using HMI data. Specifically this consisted of
significantly shorter north-south travel-time differences in the northern
hemisphere than the southern hemisphere which appears at least qualitatively similar
to the hemispheric differences between FLD frequency shifts presented here.
Our results therefore provide some qualified confirmation of this HMI-specific anomaly,
while not directly addressing differences between the results of \cite{Gizon2020} and
other HMI-based analyses \citep[e.g.][]{Zhao2013,Rajaguru2015,Chen2017}. 
Understanding the source of this anomaly is critical and, to
this end, employing FLD methods with GONG and/or MDI observations would be highly
useful.

In addition to the need to resolve this HMI-based mystery, other improvements 
to the FLD technique as applied to the study of the meridional circulation seem warranted.
Probably the most important is extending the method to model the low-$\ell$ global-mode spectra
required to infer flows in the bottom half of the convection zone. As discussed in
\S\ref{sec.powspec} the issue at hand is the blending of power ridges of adjacent radial order.
It is clear from the results shown here that employing time intervals of 8-hr duration for
the analysis is not sufficient to resolve the relevant low-$\ell$ power spectra (e.g. 
Figure~\ref{fig.pslices}). Longer duration spectra are readily obtainable
for the nominal measurements, but this is not the case for the control 
observations for which shorter duration intervals were required to minimize rotation effects.
Consequently, extending the FLD analysis deeper may require novel methods of
correcting the frequency shifts for center-to-limb artifacts.

\acknowledgments
We are grateful to Lisa Upton for providing unpublished measurements
of the surface meridional flow.
This work is supported by the 
NASA Heliophysics Division through its 
Heliophysics Supporting Research (grant 80NSSC18K0066), 
Guest Investigator (grant 80NSSC18K0068), and
Living With a Star (grant 80NSSC20K0187) programs and by the
Solar Terrestrial program of the National Science 
Foundation (grant AGS-1623844).
Y.F.'s work is supported by the National Center for Atmospheric Research, 
which is a major facility sponsored by the National Science Foundation 
under Cooperative Agreement No. 1852977.

\appendix

\section{Ridge-peak Finding Comparisons}\label{sec.append1}

We tested different peak-finding methods to extract the frequency shifts shown in
\S\ref{sec.freqshift}. In addition to the multi-ridge fitting (MRF) method
described in \S\ref{sec.freqshift}, we test two other methods. First, we employed
the centroid method used by \cite{Braun1998b}, whereby the frequency shift between
poleward and equatorward traveling waves is 
\begin{equation}
\Delta\nu(\ell,n) = \frac{\int_{\nu_1}^{\nu_2} \nu P_A(\ell,\nu) \,d\nu}{\int_{\nu_1}^{\nu_2} P_A(\ell,\nu) \,d\nu} - 
                    \frac{\int_{\nu_1}^{\nu_2} \nu P_B(\ell,\nu) \,d\nu}{\int_{\nu_1}^{\nu_2} P_B(\ell,\nu) \,d\nu},
\end{equation}
where $[\nu_1(\ell, n), \nu_2(\ell, n)]$ is a frequency window surrounding a ridge with
degree $\ell$ and radial order $n$. 
The other method involves the cross-correlation function evaluated between $P_A$ and $P_B$: 
\begin{equation}
C(\ell, n, \delta\nu) = \int_{\nu_1}^{\nu_2} P_A(\ell,\nu-\delta\nu) P_B(\ell,\nu) d\nu,
\end{equation}
where the frequency shift between poleward and equatorward traveling waves is given by
the value of $\delta\nu$ which maximizes $C$. Typically, the frequency shift is small compared
to the spacing in frequency over which the cross-correlation function is computed discretely.
Consequently, it is necessary to model and interpolate the peak of the cross-correlation 
function in some fashion. The main free parameters in these methods include
the choice of the individual frequency windows and, for the cross-correlation method, the 
means by which peak in the correlation is identified.
Estimation and removal of simple ``background'' power 
may be carried out as well. For example, \cite{Braun1998b} inferred and subtracted 
a simple linear term between high and low frequencies, in an admittedly ad-hoc fashion,
before measuring the centroid frequencies. 

In addition to comparing the results obtained using different methods applied to the observed
poleward and equatorward power spectra, we also tested each method using artificial 
power spectra. Specifically, to make these tests relevant to the applications at hand,
we use the fits of the MRF method (\S\ref{sec.freqshift}), without noise, as spectra
to test the centroid (hereafter referred to as CENT) and cross-correlation (hereafter
XCORR) methods. Not surprisingly, the MRF method applied to these noiseless spectra return the 
expected input parameters to high precision. Tests on the artificial spectra were
made with and without the cubic background term. 

\begin{figure*}
\plottwo{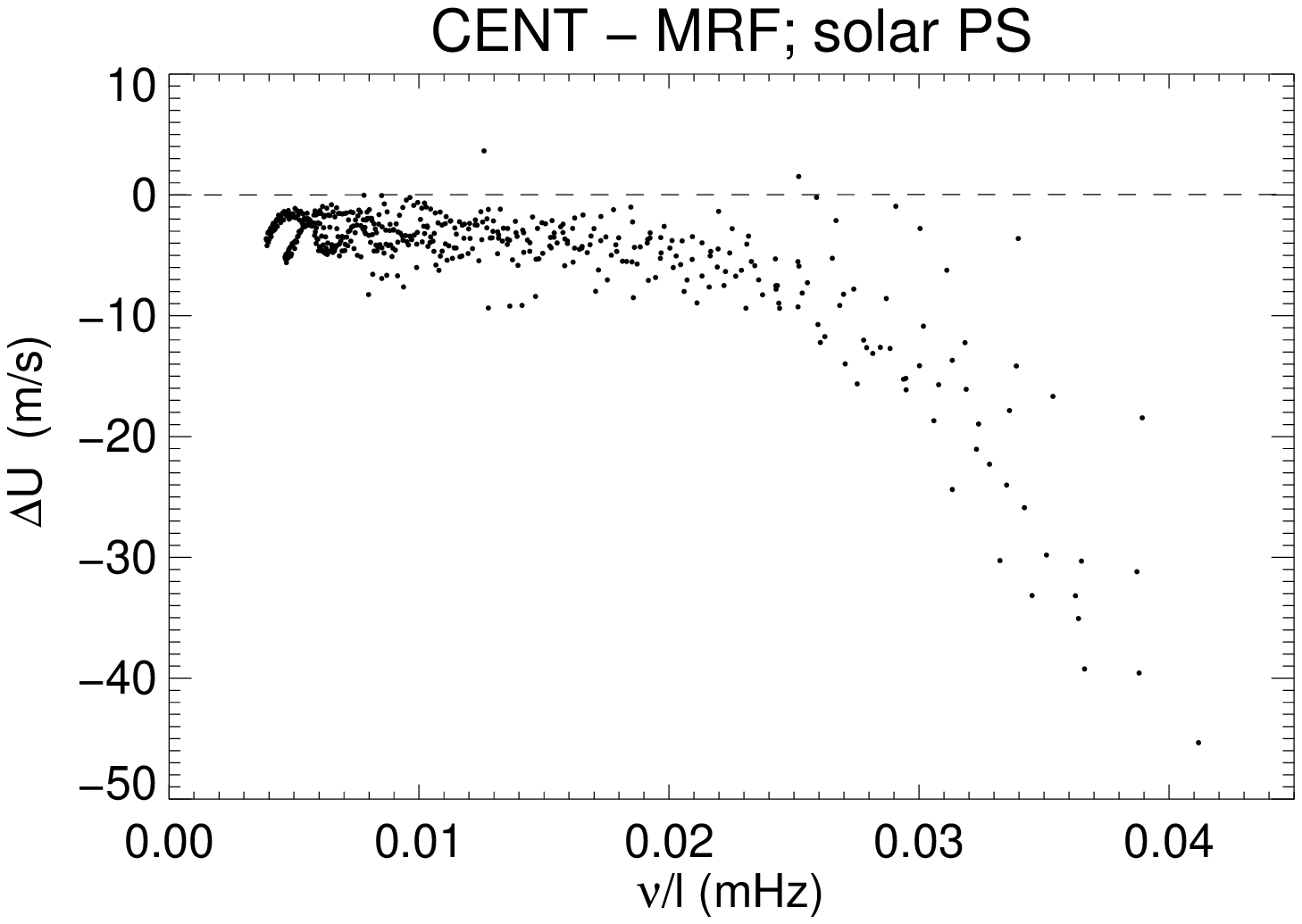}{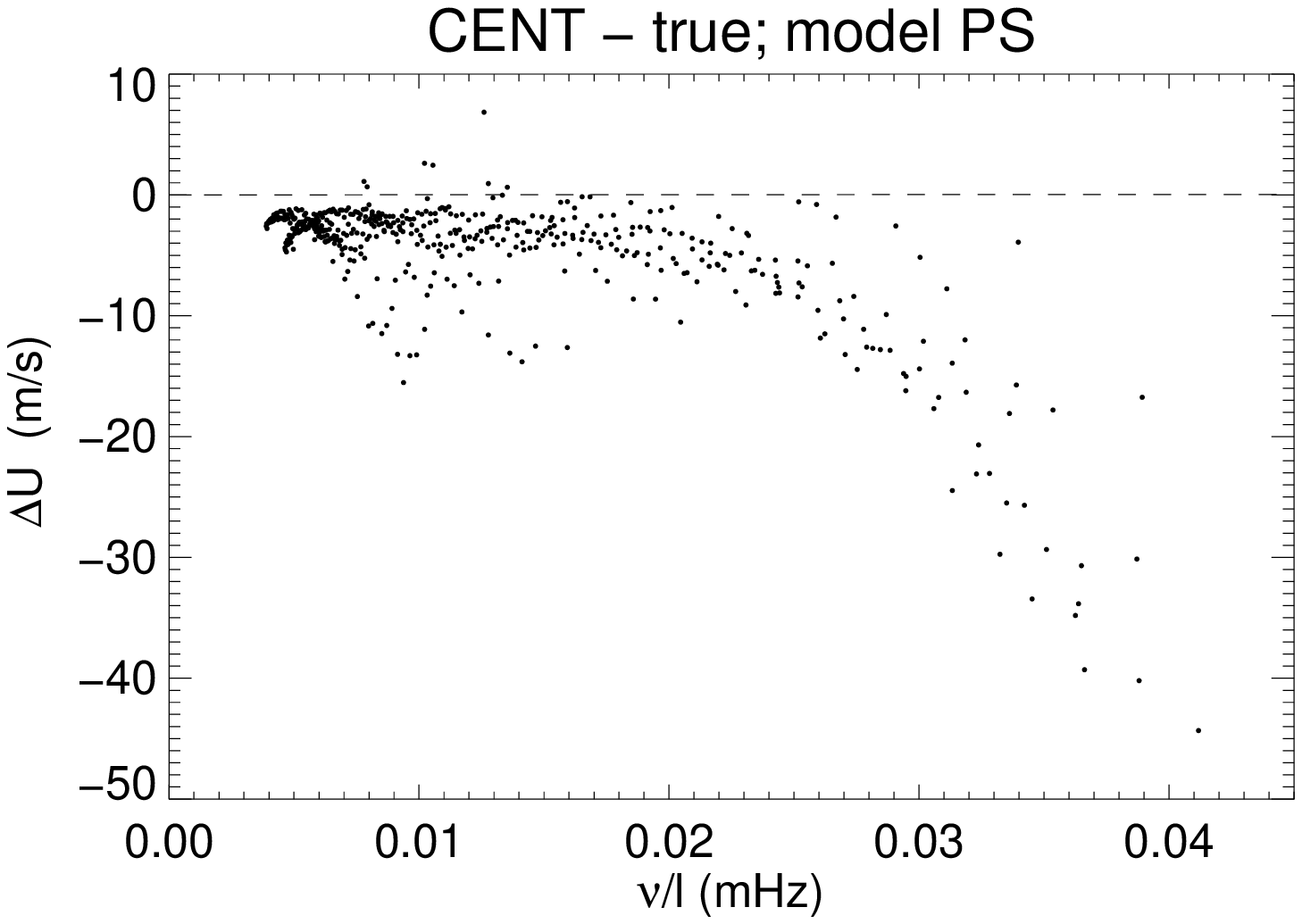}
\caption{
Tests of the centroid ridge-peaking method as described in the text. The left panel
shows the difference of the scaled frequency shifts $U$ as determined between the centroid
(CENT) method and the multi-ridge fits (MRF) as applied to the 88-month solar power
spectra in the northern hemisphere. The abscissa is the phase speed $\nu/\ell$ and the
ordinate is the frequency shift (poleward minus equatorward) obtained from the CENT method
minus the shift obtained by the MRF method. The right panel shows
the difference between the CENT frequency shifts obtained for the artificial power spectra
minus the expected (``true'') shifts in the model.
Systematic offsets between the results for the different methods, applied to solar spectra,
are observed which resemble closely the difference between the measured and expected values 
obtained from the artificial spectra.
}
\label{fig.centroid}
\end{figure*}

\begin{figure*}
\plottwo{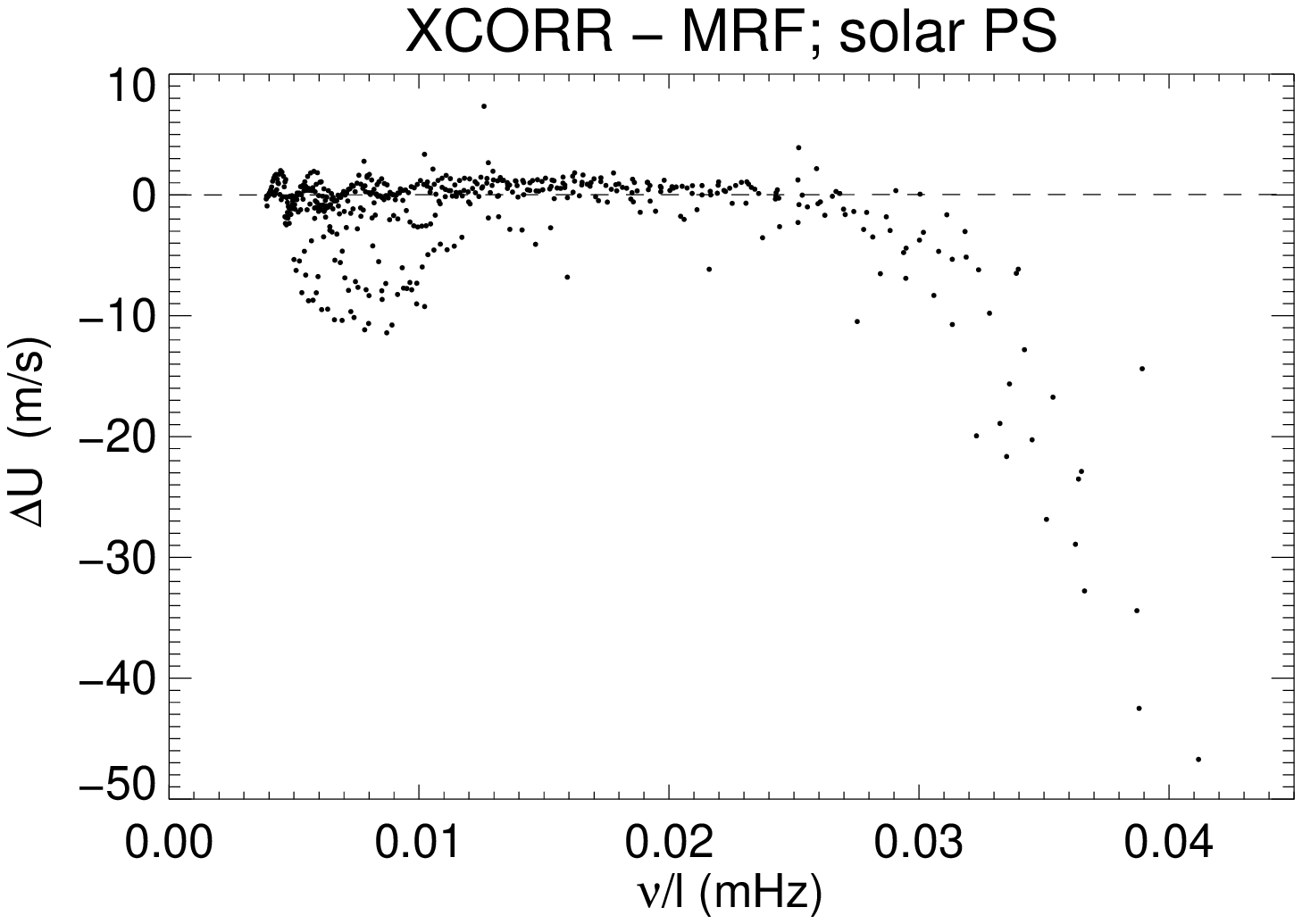}{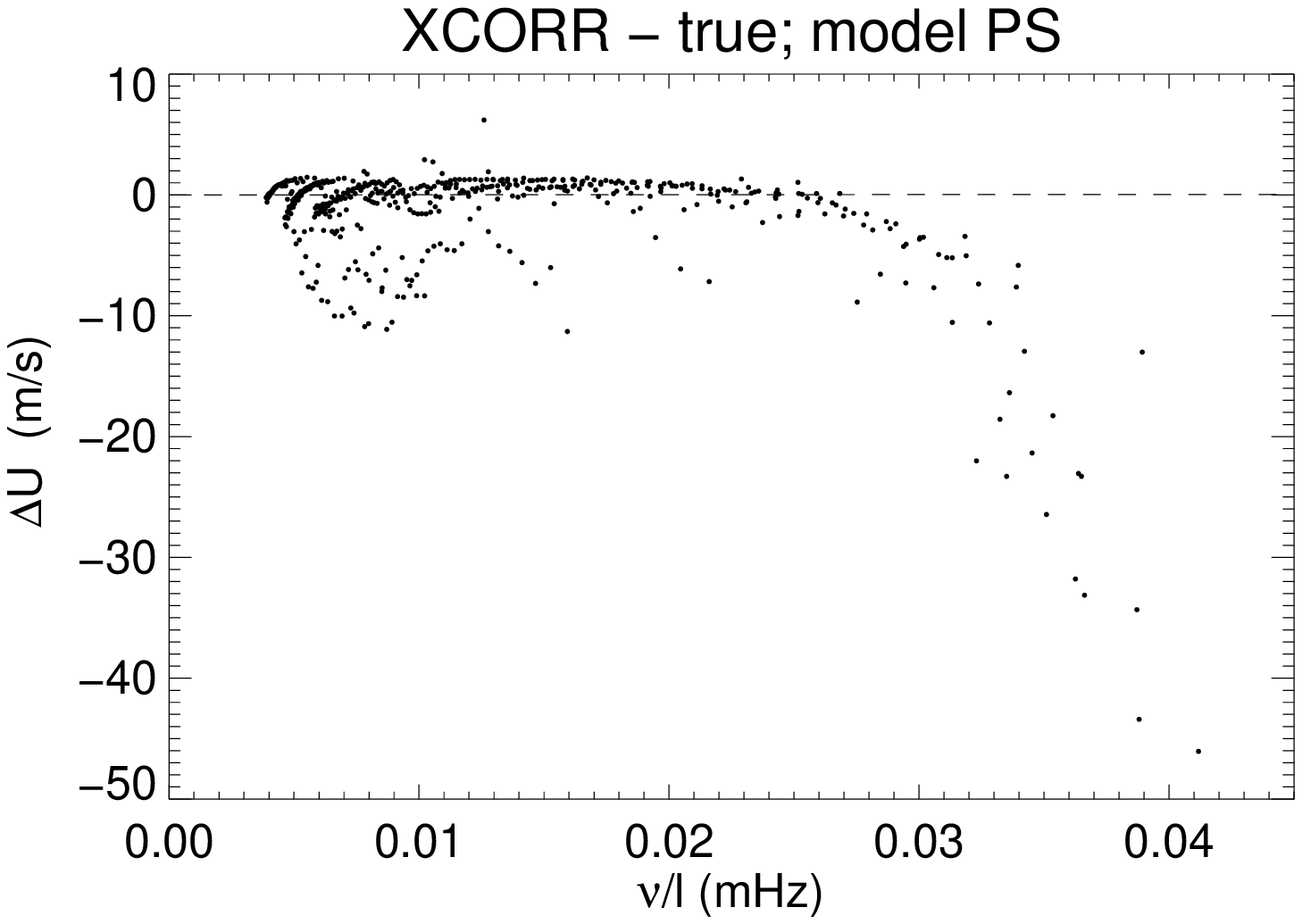}
\caption{
As Figure~\ref{fig.centroid}, but for the cross-correlation (XCORR) method. The left
panel shows the scaled frequency shift differences between the XCORR and MRF methods
applied to the solar power spectra while the right panel shows the difference between
the XCORR results, applied to the artificial spectra, from the expected values.
}
\label{fig.xcorr}
\end{figure*}

The results of these comparisons and tests are shown in Figures~\ref{fig.centroid} and
\ref{fig.xcorr}. Undeniably the most disappointing results from these tests are that
both the centroid and cross-correlation methods show significant systematic 
differences from the MRF results in a manner which is readily reproduced 
with the noiseless artificial spectra. For example, the centroid method 
consistently underestimates the shifts relative to MRF applied to
solar spectra (Figure~\ref{fig.centroid}) and likewise underestimates the true
shifts in the artificial spectra. The similarly between the left and right panels
in this figure is a reflection of how well the artificial spectra capture the
relevant properties of the actual solar power spectra in these tests. Of particular
note is the systematic bias which increases in magnitude with increasing phase speed.
The cause of this is contamination of power from nearby ridges
which becomes worse as the frequency spacing between ridges decreases with higher phase speed. 
Systematic differences at lower phase speed (i.e.\ less than 0.02 mHz) are also
observed and, in the case of the artificial spectra, are identified as contamination from the 
cubic background term in those spectra.
The interpretation of this background is uncertain in the context of the
solar spectra and may result from either an actual contribution of 
low-frequency convective signal or
an inadequacy of the MRF method to describe the properties of the mode spectra (for example,
a failure of Gaussian functions to reproduce the wings of the ridges). Whatever
its meaning, it is clear that its presence adds to the contamination problem experienced
at higher phase speeds due to the narrow ridge separation. 
Unfortunately, tests with different frequency windows, e.g.
making them narrower than the default values which span the midpoints to neighboring
ridges, produced little or no change in the bias.

Figure~\ref{fig.xcorr} shows the results obtained using the cross-correlation method.
These results were obtained by fitting the five points
bracketing the observed maximum in $C$ to the sum of coaligned quadratic and quartic functions:
\begin{equation}
\label{eqn:xcfit}
C_{\rm fit} = C_0 + C_2 (\delta\nu - \delta\nu_{\rm fit})^2 + C_4 (\delta\nu - \delta\nu_{\rm fit})^4,
\end{equation}
where $\delta\nu_{\rm fit}$ is the desired frequency shift, and $C_0$, $C_2$ and $C_4$ are 
constants.
Some improvement over the centroid method is apparent in Figure~\ref{fig.xcorr} at lower phase speeds, although
systematics still exist. At higher phase speeds we observe rapidly increasing
differences (left panel) and departures from true values (right panel) in the artificial 
spectra. Improvements (i.e. $\Delta U$ approaching zero) are observed at lower phase speeds in
the right panel when the background terms are removed, as was the case with the centroid
method. Changes in the choice of frequency windows or fitting function 
(e.g. using Gaussian functions instead of equation [\ref{eqn:xcfit}]) produced little or
no improvement in the general results shown in Figure~\ref{fig.xcorr}.

In conclusion, we find that the systematic discrepancies 
encountered with the centroid and cross-correlation
methods render them useless in preference to methods (such as MRF) which can explicitly 
account for the presence and potential contamination of neighboring ridges and possible
background (convective) power.

\section{Leakage of Meridional Flow into the Control Measurements}\label{sec.append2}

We consider the control measurements $U_{CN}$ and $U_{CS}$ and how to include 
effects of leakage of the meridional flow signal in our center-to-limb correction.
We assume that the control shifts $U_{CN}$ consist of components
due to the center-to-limb pseudo
flow, which we designate $U_{CN}^{0}$, and components due to leakage. 
We assume the latter is proportional by   
a factor $f_{\rm lk}$ to the signal produced by the meridional circulation in the nominal 
measurements. Therefore we have 
\begin{equation}
\label{eqn:leak1}
U_{CN} =  U_{CN}^{0} + f_{\rm lk} U_N^{\rm mc}
\end{equation}
where $U_{N}^{\rm mc}$ are the expected nominal frequency shifts due to 
meridional circulation alone. This are related to the raw shifts $U_N$ via a
subtraction of the true center-to-limb shifts:
\begin{equation}
\label{eqn:leak2}
U_{N}^{\rm mc} =  U_N - U_{CN}^{0}.
\end{equation}
Eliminating $U_{CN}^{0}$ from equations (\ref{eqn:leak1}) and (\ref{eqn:leak2}) yields
a modified correction operation:
\begin{equation}
\label{eqn:leak3}
U_{N}^{\rm mc} =  (1-f_{\rm lk})^{-1} (U_N - U_{CN}).
\end{equation}

We expect from a consideration of geometry (as discussed in \S\ref{sec.ctl}) that
the leakage is of opposite sign than the nominal meridional flow shifts ($f_{\rm lk} < 0$)
so the raw difference $U_N - U_{CN}$ overestimates the desired shifts $U_{N}^{\rm mc}$
and is corrected by a multiplication by a ``leakage factor'' $(1-f_{\rm lk})^{-1}$ which
is less than one.

\begin{table}[ht!]
 \begin{center}
  \caption{Leakage Factors for Different Flow Assumptions}\label{tbl.leak}
  \begin{tabular}{ccc}
   \tablewidth{0pt}
   \hline
   \hline
   {flow} & $f_{\rm lk}$ & $(1 - f_{\rm lk})^{-1}$\\
   \hline
     SF      &   -0.19$\pm 0.01$    &   0.84$\pm 0.01$\\         
   \hline
    obs (N)  &   $\leq -0.13$       &  $\leq 0.89$    \\   
    obs (S)  &   $\leq -0.15$       &  $\leq 0.87$    \\
   \hline
      a      &   -0.28              &   0.78          \\
      b      &   -0.26              &   0.79          \\
      c      &   -0.24              &   0.81          \\
      d      &   -0.19              &   0.84          \\
      e      &   -0.17              &   0.85          \\        
      f      &   -0.17              &   0.85          \\      
   \hline
  \end{tabular}
 \end{center}
\end{table}

To constrain the leakage factor, we first 
consider the shallowest modes (i.e. small phase speeds) and
estimate the leakage caused by
surface values of the meridional flows. 
For this purpose, we acquired latitudinal 
profiles (Upton, private communication) of the 
surface meridional flows averaged over each Carrington rotation over our 88-month
interval and determined from feature tracking \citep{Hathaway2010,Rightmire-Upton2012}.
The profiles take the form of coefficients to polynomial fits to 
the flow up to fifth order in $\sin(\lambda)$. For each profile, we perform
a coordinate transformation on the vector
flow field as viewed in the nominal coordinate system (aligned to the rotation axis) to
determine its ``inward/outward'' component in the coordinate system aligned to the 
pseudo poles used in the control measurements. The leakage
ratio $f_{\rm lk}$ is then given by the spatial average of this component over
the observed window function $W$ divided by the average of the north/south
component in the original coordinate system over the same window. Over the relevant time frame,
we obtain the mean and standard deviations of $f_{\rm lk}$ as well as 
the factor $(1 - f_{\rm lk})^{-1}$. These values, labeled as flow ``SF,''  are 
listed in the first line of Table~\ref{tbl.leak}.

Prior observations have suggested that the center-to-limb pseudo flow trends towards zero 
amplitude as the phase speed (or equivalently, skip-distance in time-distance
measurements) decreases \citep[e.g.][]{Greer2013,Chen2017}. Our own observations
(e.g. Figure~\ref{fig.ucn})
uniquely show negative values of the control shifts $U_{CN}$ and $U_{CS}$ for the
$f$ and $p_1$ modes at the smallest phase speeds 
which plausibly arise to the leakage discussed here. Under the assumption
that $U_{CN}^0$ approaches zero for the shallowest modes, we assess the
ratio of $U_{CN}/U_N$ and $U_{CS}/U_S$, as averaged over phase speeds $< .004$ mHz.
These values,
as listed in Table~\ref{tbl.leak} and labeled ``obs (N)'' and ``obs (S)'' respectively.
represent upper limits to $f_{\rm lk}$, thus allowing some small positive 
center-to-limb contribution $U_{CN}^0$. but are consistent with the expectation from
the surface flows ``SF.''

\begin{figure}
\plotone{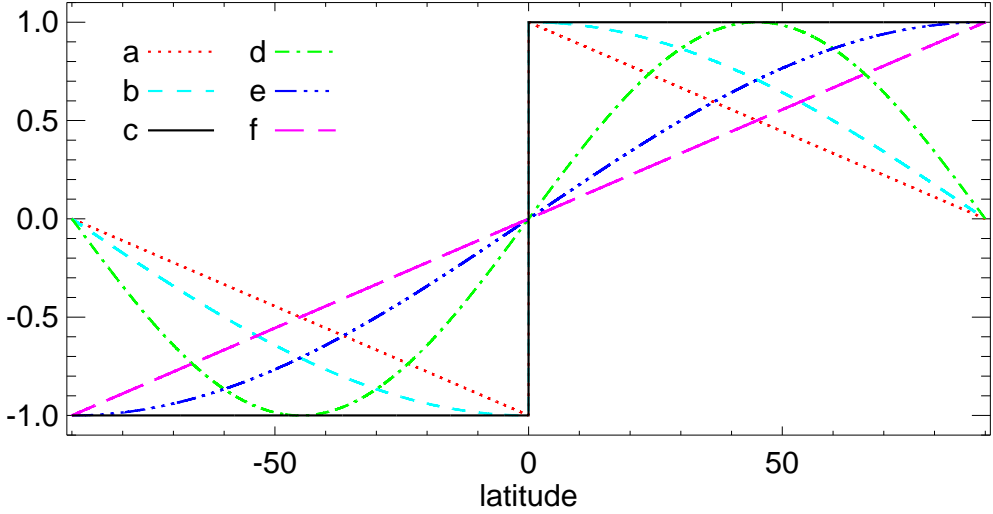}
\caption{
Simple functions used to estimate the leakage of the meridional flow signal into
the control shifts. The labels refer to the functions used to produce
the leakage factors listed in Table~\ref{tbl.leak}.
}
\label{fig.flowfuncs}
\end{figure}

The remaining task is to estimate the leakage expected for the deeper measurements.
It is tempting to consider the possibility that the leakage is invariant (or at least
sufficiently so) over our range of frequency-shift measurements, which enables us to multiply
all shifts with a simple constant in the correction (\ref{eqn:leak3}). 
To this end, it is worth considering what conditions might produce large variations
in the leakage with depth. It is clear from the geometry of the control measurements
that the range in $\lambda$ sampled by the control annulus extends to the equator, unlike
the nominal measurements which are cut off for $|\lambda| < 20^\circ$. Thus, a redistribution
of the flow in latitude with increasing depth needs to be considered.
Using the same coordinate transformation and integration described above,
We compute the variation of the factor $(1 - f_{\rm lk})^{-1}$ among different toy models 
of the $\lambda$ dependence. Some assumptions which help to constrain possible profiles
include: (1) the meridional circulation remains nearly antisymmetric about the equator, and 
(2) its variation with latitude is smooth and maintains the same sign (poleward or equatorward)
over each hemisphere. Using the simple functions illustrated
in Figure~\ref{fig.flowfuncs} we obtain the  results listed in Table~\ref{tbl.leak}.
We note that a flow given by a multiplication of the functions shown 
in Figure~\ref{fig.flowfuncs} by a constant with either sign will produce the same leakage. 
Thus, these results are relevant equally for poleward and equatorward (e.g. return) flows.
The smallest (largest) values of the leakage factor occur when the relative contributions
of the flow are concentrated more at the lower (higher) latitudes. However,
variations of the leakage parameter remain comfortably modest over all of the cases
considered.

On the basis of these results, we assume for our inferences here that the leakage
factor has a nominal value of 0.84. However, in evaluating the results 
inferred in \S\ref{sec.forward},
we consider the possibility that this factor varies with mode depth 
within a conservative range of 0.75 to 0.90, which represents a net uncertainty of about
20\%.

\section{Sensitivity Functions}\label{sec.append3}

\begin{figure*}
\plottwo{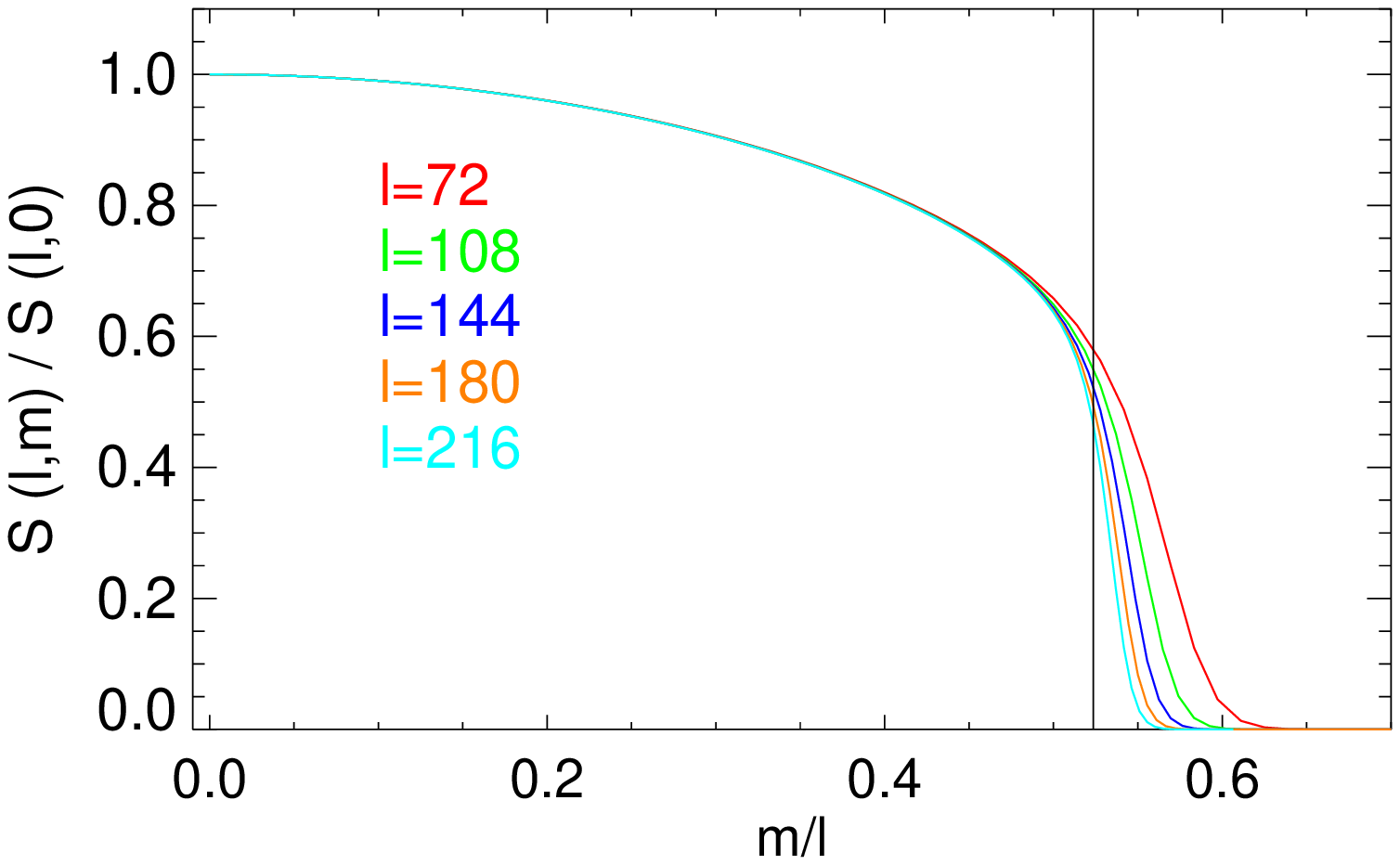}{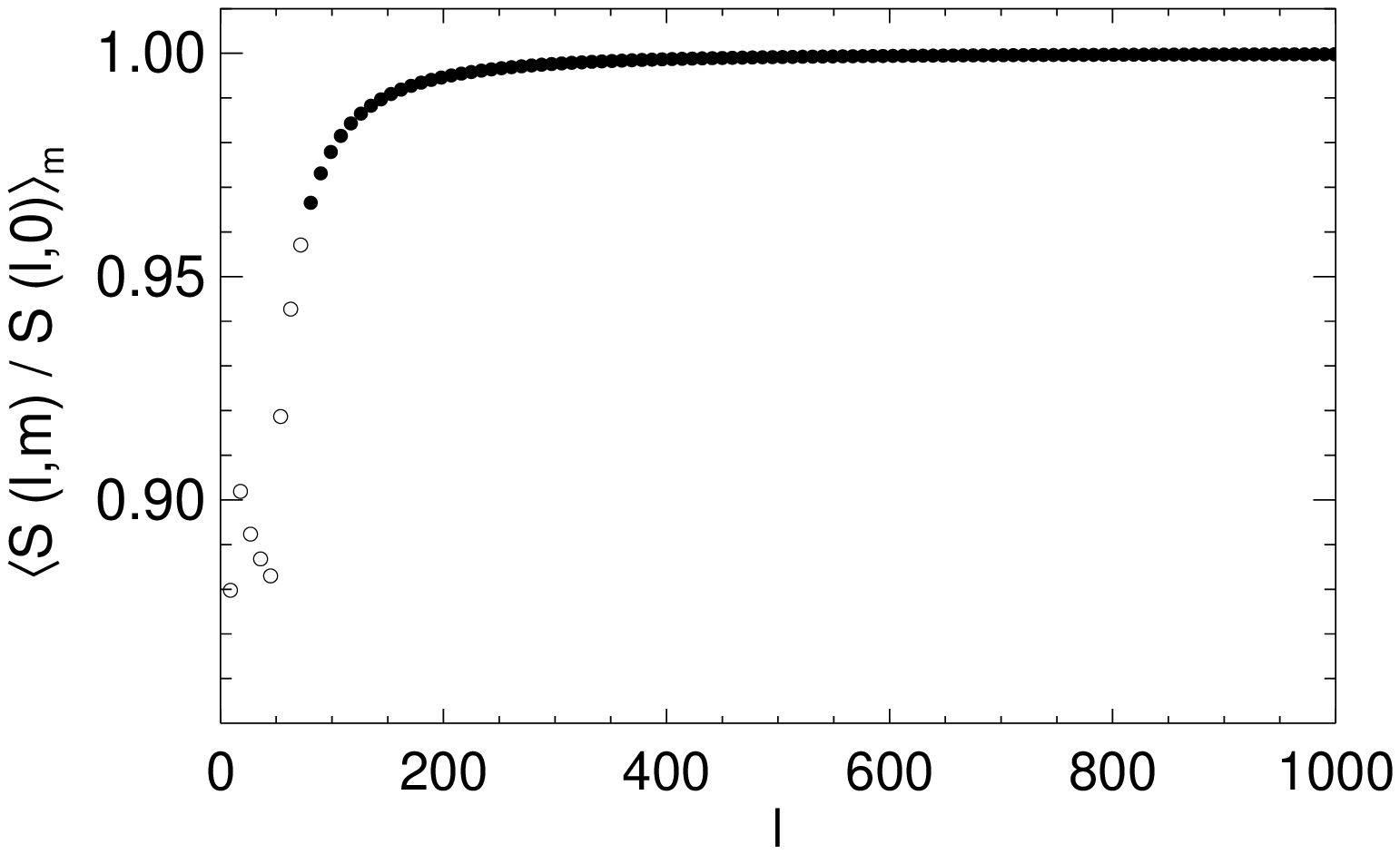}
\caption{
The sensitivity of the frequency shifts to a meridional directed flow for a mode with
degree $\ell$ and azimuthal order $|m|$, relative to a mode with the same degree but $m=0$.
The left panel shows the dependence of the function defined by equation (\ref{eqn:msens}) 
on $|m|/\ell$. The different curves represent different values of $\ell$ as indicated. 
The vertical line
represents $|m|/\ell = 0.5236$ which is the limit imposed by equation (\ref{eqn:impact}).
The right panel shows the average of this function over the azimuthal orders for which
power spectra were summed (see equation \ref{eqn:powersum}). The filled circles indicate
values of mode degree which were used in the forward modeling (\S\ref{sec.forward}), while
open circles indicate modes not used.
}
\label{fig.mdepend}
\end{figure*}

The sensitivity of the frequency shift $\Delta\nu(\ell,m,n)$ to an interior flow
$\mathbf{v}(\theta,r)$, which is steady over the time interval $T$ of the analysis 
and smooth in $\theta$ and $r$, 
can be derived by a perturbation analysis of the wave equation in the presence
of the flow \citep{Gough1983}. In spherical coordinates, one obtains: 
\begin{multline}
\label{eqn:kernel1}
\Delta\nu(\ell,m,n) =  \frac{2}{\pi^2}\frac{(\ell+m)!}{(\ell-m)!} 
\left(\int_{\theta_1}^{\theta_2} \left[ (P_{\ell}^m (\cos\theta))^2 + \frac{4}{\pi^2}((Q_{\ell}^m (\cos\theta))^2\right] \sin\theta d\theta \right)^{-1}\\
\times \frac{ \displaystyle\int_0^{R_\odot} \left[ \displaystyle\int_{\theta_1}^{\theta_2} v_{\theta}(\theta,r) d\theta \right] \left[ \xi_{\ell n}^2 +\ell(\ell+1)\eta_{\ell n}^2 \right] \rho r dr}
{ \displaystyle\int_0^{R_\odot} \left[ \xi_{\ell n}^2 +\ell(\ell+1)\eta_{\ell n}^2 \right] \rho r^2 dr}
\end{multline}
where $\rho$ is the density, and $\xi_{\ell n}$ and $\eta_{\ell n}$ are related to the components of the eigenmodes
\begin{equation}
\label{eqn:eigenm}
\boldsymbol{\xi}_{\pm}(\ell,m,n) = \left[ \xi_{\ell n}(r), \eta_{\ell n}(r) \frac{\partial}{\partial\theta}, \eta_{\ell n}(r) \frac{1}{\sin\theta}\frac{\partial}{\partial\phi}\right] 
\left[ P_{\ell}^m (\cos \theta) \pm \frac{2i}{\pi} Q_{\ell}^m (\cos \theta) \right] e^{im\theta},
\end{equation}
``$\pm$'' distinguishes between poleward and equatorward traveling wave modes, and
$v_{\theta}$ is the component of the flow $\mathbf{v}$ in the meridional direction $\theta$.
We can rewrite equation (\ref{eqn:kernel1}) in terms of the scaled frequency shift $U$:
\begin{equation}
\label{eqn:kernel2}
U (\ell, m, n) = \left(\frac{S(\ell,m)}{S(0,m)}\right) 
         \frac{ {R_\odot} \displaystyle\int_0^{R_\odot} \frac{\left\langle v_\theta \right\rangle}{r} K_{\ell n}(r) dr}{\displaystyle\int_0^{R_\odot} K_{\ell n}(r) dr},
\end{equation}
where $\langle v_{\theta} \rangle$ 
is the average of the meridional flow component 
(equation \ref{eqn:vthet}),
$\Delta \theta \equiv \theta_2 - \theta_1$, and the  kernel $K_{\ell n}$ is given by the
mode kinetic energy density:
\begin{equation}
\label{eqn:kernel3}
K_{\ell n} = \left[ \xi_{\ell n}^2 +\ell(\ell+1)\eta_{\ell n}^2 \right] \rho r^2.
\end{equation}
The function $S(\ell,m)$ contains the terms on the right hand side of equation (\ref{eqn:kernel1})
which depend on the azimuthal order $m$:
\begin{equation}
\label{eqn:msens}
S(\ell,m) = \frac{(\ell+m)!}{(\ell-m)!} \left(\int_{\theta_1}^{\theta_2} \left[ (P_{\ell}^m (\cos\theta))^2 + \frac{4}{\pi^2}((Q_{\ell}^m (\cos\theta))^2\right] \sin\theta d\theta \right)^{-1},
\end{equation}
and $S(\ell, 0) \approx (\pi\ell/2\Delta\theta)$ to a good approximation.
The ratio $S(\ell,m)/S(\ell, 0)$ gives the sensitivity of the frequency shift to 
azimuthal order, relative to $m=0$, and is shown in the left panel of Figure~\ref{fig.mdepend}.
The ratio reduces to a single-valued function of $|m|/\ell$, sometimes referred to as
the ``impact parameter'' in scattering theory, except when this parameter approaches and exceeds
the limit imposed by equation (\ref{eqn:impact}). The right panel of Figure~\ref{fig.mdepend}
shows the average of this ratio over the azimuthal orders for which the power spectra
were summed in this work. For the modes used in the forward modeling $(81 \leq \ell \leq 999)$
the values of this ratio are between about 0.97 and 1.0. In the modeling (\S\ref{sec.forward})
we compare the frequency shifts from the $m$-summed power spectra with predictions made
using equation (\ref{eqn:kernel2}) with the ratio replaced by the averages
shown in the right panel of Figure~\ref{fig.mdepend}.

\bibliographystyle{/export/home/dbraun/Macros/apj}
\bibliography{/export/home/dbraun/Macros/db}

\end{document}